\documentclass[pdflatex,referee,sn-basic]{sn-jnl}

\usepackage{graphicx}%
\usepackage{multirow}%
\usepackage{amsmath,amssymb,amsfonts}%
\usepackage{amsthm}%
\usepackage{mathrsfs}%
\usepackage[title]{appendix}%
\usepackage{xcolor}%
\usepackage{textcomp}%
\usepackage{manyfoot}%
\usepackage{booktabs}%
\usepackage{algorithm}%
\usepackage{algorithmicx}%
\usepackage{algpseudocode}%
\usepackage{listings}%

\usepackage{natbib}
\usepackage{bm}
\usepackage{array} % fixes undefined \arraybackslash error
\usepackage{comment}
\usepackage{float}
\usepackage{ragged2e}

\DeclareMathOperator*{\argmin}{arg\,min}

\begin{document}

% \title[Heterogeneity in Integration and Prediction (HIP)]{HIP: a method for high-dimensional multi-view data integration and prediction accounting for subgroup heterogeneity}
\title[Heterogeneity in Integration and Prediction (HIP)]{Accounting for data heterogeneity in integrative analysis and prediction methods: An application to Chronic Obstructive Pulmonary Disease}

%Accounting for data heterogeneity in integrative analysis and prediction methods: An application to Chronic Obstructive Pulmonary Disease (COPD)


\author[1]{\fnm{Jessica} \sur{Butts}}\nomail
\author[2]{\fnm{Christine} \sur{Wendt} \dgr{MD}}\nomail
\author[3]{\fnm{Russel P.} \sur{Bowler} \dgr{MD, PhD}}\nomail
\author[4]{\fnm{Craig P.} \sur{Hersh} \dgr{MD}}\nomail
\author[5]{\fnm{Qi} \sur{Long} \dgr{PhD}}\nomail
\author[1]{\fnm{Lynn} \sur{Eberly} \dgr{PhD}}\nomail
\author*[1]{\fnm{Sandra E.} \sur{Safo} \dgr{PhD}}\email{ssafo@umn.edu}

\affil*[1]{\orgdiv{Division of Biostatistics}, \orgname{University of Minnesota}, \orgaddress{\city{Minneapolis}, \state{MN}, \country{USA}}}
\affil[2]{\orgdiv{Division of Pulmonary, Allergy and Critical Care}, \orgname{University of Minnesota}, \orgaddress{\city{Minneapolis}, \state{MN}, \country{USA}}}
\affil[3]{\orgdiv{Division of Pulmonary, Critical Care and Sleep Medicine, Department of Medicine}, \orgname{National Jewish Health}, \orgaddress{\city{Denver}, \state{CO}, \country{USA}}}
\affil[4]{\orgdiv{Channing Division of Network Medicine, Brigham and Women's Hospital}, \orgname{Harvard Medical School}, \orgaddress{\city{Boston}, \state{MA}, \country{USA}}}
\affil[5]{\orgdiv{Department of Biostatistics, Epidemiology and Informatics,
 Perelman School of Medicine}, \orgname{University of Pennsylvania}, \orgaddress{\city{Philadelphia}, \state{PA}, \country{USA}}}

%The Abstract should not exceed 350 words. Please minimize the use of abbreviations and do not cite references in the abstract. The abstract must include the following separate sections:

% Background: the context and purpose of the study
\abstract{\textbf{Background:} Epidemiologic and genetic studies in chronic obstructive pulmonary disease (COPD) and many complex diseases suggest subgroup disparities (e.g., by sex) in disease course and patient outcomes. We consider this from the standpoint of integrative analysis where we combine information from different views (e.g., genomics, proteomics, clinical data). Existing integrative analysis methods ignore the heterogeneity in subgroups, and stacking the views and accounting for subgroup heterogeneity does not model the association among the views. We propose HIP (Heterogeneity in Integration and Prediction), a statistical approach for joint association and prediction that leverages the strengths in each view to identify molecular signatures that are shared by and specific to males and females and that contribute to the variation in COPD, measured by airway wall thickness.\\
% 
% Results: the main findings
\textbf{Results:} Our COPD findings have identified proteins, genes, and pathways that are common across and specific to males and females, some of which have been implicated in COPD, while others could lead to new insights into sex differences in COPD mechanisms. \\
% 
% Conclusions: a brief summary and potential implications
\textbf{Conclusions:} HIP  accounts for subgroup heterogeneity in multi-view data, ranks variables based on importance, is applicable to univariate or multivariate continuous outcomes, and incorporates covariate adjustment. With the efficient algorithms implemented using PyTorch, this method has many potential scientific applications.}


\maketitle

%------------------------------------------------------------------------------
\section{Background}\label{sec:intro}

% updated 

Chronic obstructive pulmonary disease (COPD) is a chronic progressive disease affecting more than 16 million adults,  presenting a substantial and increasing economic and social burden \citep{wheaton2015employment}; COPD was projected to  cost the U.S. economy about $\$49$ billion in 2020 \citep{guarascio2013clinical}. Although tobacco smoking is the leading environmental risk factor for COPD, even in heavy smokers fewer than 50\% develop COPD \citep{GOLD:2020}; genetics \citep{hardin2014chronic}, environmental exposures \citep{hu2010risk}, inflammation \citep{pauwels2001global} and other factors \citep{chung2008multifaceted} predispose individuals to develop COPD. The Genetic Epidemiology of COPD (COPDGene) Study \citep{regan2011genetic} is one of the largest studies to investigate the underlying genetic factors of COPD to understand why certain smokers develop COPD while others do not. While  many genomic studies have successfully identified multiple genetic variants for COPD susceptibility, most identified genetic variants do not reside in protein-coding regions \citep{silverman2018applying} making it difficult to interpret their function. Genomics data used in combination with other omics (e.g., proteomics) and known risk factors show promise in identifying multifaceted features that can enhance our understanding of mechanisms of COPD susceptibility. 

Epidemiologic and genetic studies suggest subgroup (e.g., sex) disparities exist for many complex diseases. Subgroups of a population can present similar symptoms but have different clinical courses and respond to therapy differently. By determining factors predictive of an outcome for each subgroup, we can better personalize treatments to improve patient outcomes. Research suggests sex disparities exist in COPD mechanisms \citep{barnes2016sex}. A meta-analysis of 11 studies showed that female smokers, even if smoking fewer cigarettes, had a faster annual decline in forced expiratory volume in one second (FEV$_{1}$) \citep{gan2006female}. A study using COPDGene data found women smokers tended to have higher airway wall thickness (AWT) compared to male smokers \citep{kim2011gender}, likely explaining some of the sex differences in the prevalence of COPD. Women with severe COPD may be at higher risk for hospitalization and death \citep{prescott1997gender}. These studies primarily used data from one source, so combining data from multiple sources has the potential to reveal new insights into sex differences in COPD mechanisms. Motivated by the crucial scientific need to understand sex differences in COPD, we leverage the strengths from multiple data views from the COPDGene Study %\citep{regan2011genetic} 
to identify genes and proteins \textit{common among} and \textit{specific to} males and females contributing to variation in AWT. 

Existing methods for integrating data from multiple views are inadequate for our problem as they do not account for subgroup heterogeneity. In particular, one-step methods have been proposed for joint association of data from multiple views and simultaneous prediction of an outcome \citep{SIDA,BIPNet,luo2016canonical}. 
To do this, one would build a separate integrative analysis model for each subgroup to determine the important multidimensional variables that are associated and predictive of the outcome for each subgroup. 
While this approach is intuitive, it is limited by the sample size for each subgroup and does not pool information across subgroups making estimation challenging. %less precise and subject to potential problems with convergence.
This is especially true for high-dimensional data settings where the number of variables is larger than the sample size for each subgroup. Another approach that makes use of samples in all subgroups is to apply these one-step methods on the combined subgroup data, but this precludes us from examining whether such heterogeneity exists. 

The need to account for subgroup heterogeneity has been recognized and studied in the case where there is only one data view. \cite{jointlasso} propose the Joint Lasso to jointly estimate regression coefficients for different subgroups while allowing for the identification of subgroup-specific features, and also encouraging similarity between subgroup-specific coefficients. In \cite{li2014meta}, the authors proposed the meta lasso for feature selection for different studies (in our application, subgroups) that incorporates a hierarchical penalty to borrow strength across different studies, while allowing for feature selection flexibility. The goal of the meta lasso is to combine data sets with the same variables measured on distinct subjects from separate studies to improve variable selection across all data sets when considering a binary outcome; it is not specifically designed to account for subgroup heterogeneity and does not consider multiple data views obtained on the same set of subjects. To use these existing methods, one would stack the different data views for each subgroup; this approach assumes the many variables across the data views are independent and ignores the overall dependency structure among the different views. 

We make three main contributions in this article. First, we propose integrative analysis and prediction methods that account for subgroup heterogeneity and are appropriate for our motivating data by modifying the hierarchical penalty proposed in \cite{li2014meta} to improve power for identifying common and subgroup-specific features. Second, the methods we propose, called HIP (Heterogeneity in Integration and Prediction), allow for one or more continuous outcomes, and can force specified covariates into the model giving HIP more flexibility than current methods. Third, we develop computationally efficient algorithms using PyTorch \citep{PyTorch}. We apply the methods to our motivating data from the COPDGene Study to identify genes and proteins common across and specific to males and females and associated with AWT. We then explore enriched pathways and the ability of these omics biomarkers to predict AWT beyond some established COPD risk factors. 

The remainder of the paper is structured as follows. In Section \ref{sec:meth}, we present the proposed methods (HIP). In Section \ref{sec:alg}, we describe the implementation of HIP. In Section \ref{sec:sim}, we conduct simulation studies to assess the performance of HIP in comparison with existing methods. In Section \ref{sec:real}, we apply HIP to data from the COPDGene Study. We conclude with some brief discussion in Section \ref{sec:conc}.

%-----------------------------------------------------------------------
\section{Methods}\label{sec:meth}

% updated
\subsection{Notation and Problem}
Suppose we have $D$ views (e.g., genomics, proteomics, clinical) with $p_d$ variables measured on the same $N$ subjects. Each view has $s =1,\ldots,S$ subgroups known a priori, each with sample size $n_s$ where $N=\sum_{s=1}^{S}n_s$. For subgroup $s$ and view $d$, $\bm X^{d,s} \in \Re^{n_s \times p_d}$ represents the data matrix. Assume we also have outcome data for each subgroup. For a continuous outcome(s) (e.g., AWT), we have matrix $\bm Y^s \in \Re^{n_s \times q}$ where $q$ is the number of outcomes. Our primary goal is to perform integrative analysis that considers the overall dependency structure among views, predicts an outcome, incorporates feature ranking, and accounts for subgroup heterogeneity to identify common and subgroup-specific variables contributing to variation in the outcome.  

% updated
\subsection{Integration of multi-view data}
To relate the views within each subgroup, we assume there are subgroup-specific scores ($\bm Z^{s}$) that drive the dependency structure among the views. Then, each view is written as the product of the subgroup-specific scores $\bm Z^s \in R^{n_s \times K}$ and a matrix of view and subgroup-specific loadings $\bm B^{d,s} \in R^{p_d \times K}$ plus a matrix of errors: $\bm X^{d,s} = \bm Z^s {\bm B^{d,s}}^T + \bm E^{d,s}$. Here, $K$ is the number of components used to approximate each view. The $\bm Z^s$ incorporates the correlation across the $D$ views for subgroup $s$, and $\bm E^{d,s}$ accounts for the remaining variability unique to view $d$ for subgroup $s$. In optimizing $\bm Z^s$ and ${\bm B^{d,s}}$ we want to minimize the error in reconstructing $\bm X^{d,s}$, i.e., $\bm E^{d,s}$, via the loss function $F(\bm X^{d,s}, \bm Z^s, \bm B^{d,s})= \|\bm X^{d,s} - \bm Z^s {\bm B^{d,s}}^T\|_F^2$. For a random matrix $\bm A$, $\|\bm A\|_F^2$ is the square of the Frobenius norm defined as trace($\bm A^{T} \bm A$). 

The decomposition of $\bm X^{d,s}$ in our approach is motivated by a principal components framework rather than a factor analytic framework as we do not impose any distribution on $\bm Z^s$ or $\bm E^{d,s}$. Typically we would require ${\bm B^{d,s}}^T {\bm B}^{d,s} = \bm I$, and ${\bm Z^s}^T {\bm Z^s} = \bm I$ for uniqueness, but we do not require these constraints because we are interested in whether a variable's estimated coefficients in $\bm B^{d,s}$ are zero or not. Since we propose to use a penalty that encourages row-sparsity, we preserve the sparsity pattern in $\bm B^{d,s}$ over matrix multiplication. Further, we only use $\bm Z^s$ to predict the clinical outcome and not to make inference on the estimates in $\bm Z^s$.

\subsection{Hierarchical Penalty for Common and Subgroup-Specific Feature Ranking}\label{subsec:HIPpenalty}

% updated
A main goal in this paper is to identify common and subgroup-specific features associated with an outcome. Based on the hierarchical reparameterization proposed in \cite{li2014meta}, we decompose $\bm B^{d,s}$ as the element-wise product of $\bm G^d$ and $\bm \Xi^{d,s}$ i.e., $\bm B^{d,s} = \bm G^d \cdot \bm \Xi^{d,s}$ for $d=1,\ldots,D$ and $s=1,\ldots,S$ to estimate effects that are common across subgroups using $\bm G^d \in \Re^{p_d \times K}$ while also allowing for heterogeneity in the subgroups through $\bm \Xi^{d,s} \in \Re^{p_d \times K}$. If there is no heterogeneity, then $\bm \Xi^{d,s}$ is a matrix of ones for all $s$, and $\bm G^d = \bm B^d$, i.e., the view-specific loadings are the same for all subgroups. In estimating $\bm G^d$, we borrow strength across subgroups for increased power. In this reparameterization, exact values of $\bm G^d$ and $\bm \Xi^{d,s}$ are not identifiable, but also are  not directly needed for variable ranking.

% updated
We use regularization to induce sparsity by adding the block $l_1/l_2$ penalty on $\bm G^d$ and $\bm \Xi^{d,s}$: 
\begin{equation}
    \sum_{s=1}^S \mathcal{J}(\bm B^{d,s}) = \lambda_G \gamma_d \sum_{l=1}^{p_d} \| {\bm g_l^d}\|_2 + \lambda_\xi \gamma_d  \sum_{s=1}^S\sum_{l=1}^{p_d} \| \bm \xi_l^{d,s}\|_2 \label{penalty}.
\end{equation}
Here, $\bm g_l^d$ and $\bm \xi_l^{d,s}$ are the $l$th rows in $\bm G^d$ and $\bm \Xi^{d,s}$ respectively and are each length $K$. By imposing the block $l_1/l_2$ penalty on the rows of $\bm G^d$ and $\bm \Xi^{d,s}$, the $K$ components are considered as a group, encouraging variables to be selected in all $K$ components or not to be selected. This is desirable because the selection of variables is not component-dependent and thus appropriate for variable screening. This differs from the original hierarchical penalty reparameterization proposed in \cite{li2014meta} which imposes an $l_1$ penalty. Both $\lambda_G$ and $\lambda_\xi$ are tuning parameters controlling feature selection. Specifically, $\lambda_G$ controls feature selection for all subgroups combined and encourages removal of variables that are not important for all $S$ subgroups. Also, $\lambda_\xi$ encourages feature selection for each subgroup. Further details on the selection of $\lambda_G$ and $\lambda_\xi$ are given in Section \ref{tuning}. The $\gamma_d$ is a user-specified indicator for whether the view should be penalized. This is to allow some views, such as a set of clinical covariates, to be forced into the model to guide the selection of other important variables, which can result in better prediction of the outcome.

% updated
\subsection{Relating Shared Scores to Clinical Outcome(s)}
Besides identifying the common and subgroup-specific features, we aim to predict a clinical outcome while allowing for heterogeneity in effects based on the subgroup and multi-view data. We assume the outcome is related to the views only through the shared scores, i.e. $Z^s$, for each subgroup. This allows us to couple the problem of associations among different views and predicting an outcome. We relate the outcome to $\bm Z^s $ by minimizing a loss function $\sum_{s=1}^S F(\bm Y^s, \bm Z^s, \bm \Theta, \beta_0)$. For continuous outcome(s), $F(\bm Y^s, \bm Z^s, \bm \Theta, \beta_0) = || \bm Y^s - (\beta_0 + \bm Z^s \bm \Theta)||_F^2$, where $\bm \Theta \in \Re^{K \times q}$ are regression coefficients. 

We can impose the constraint that the columns of $\bm Z^s$ are uncorrelated ($\bm Z^{s^T}\bm Z^s=\bm I$) so each of the $K$ components provides unique information. Of note, our goal is not to interpret the coefficients $\bm \Theta$ as in regression analysis; our goal is to rank the variables corresponding to those coefficients. In our applications, we standardize each column of $\bm Y^s$ to have mean $0$ and variance $1$ at the subgroup level, but this is not necessary because of the estimation of the intercept $\beta_0$. 

Our proposal to model one or more continuous outcomes in a one-step integrative analysis model that that provides predictions based on rank-selected features, allowing for subgroup heterogeneity in those features, is novel and will be of use in many scientific applications.

% upated
\subsection{Joint Model for Integration and Prediction}\label{sec:jointmodel}

Typical integration and prediction methods follow two steps. First, the subgroup-specific scores $\bm Z^s$ and view and subgroup-specific loadings $\bm B^{d,s}$ (hence common and subgroup-specific variables, $\bm G^d$ and $\bm \Xi^{d,s}$) are learned. Second, the learned $\bm Z^s$ are associated with the outcome in a regression model. Since these steps are independent, the common and subgroup-specific variables identified may not be meaningfully connected to the clinical outcome. To overcome this limitation, we use the outcome to guide the selection of the common and subgroup-specific variables in a joint model. Thus, we use HIP to estimate the following: view and subgroup-specific loadings $\bm B^{d,s}$ ($\bm G^d$, the common variables, and  $\bm \Xi^{d,s}$, the subgroup-specific variables), the subgroup-specific scores shared across views ($\bm Z^s$), and the regression estimates ($\bm \Theta$, $\beta_0$).
To obtain these estimates, we combine the outcome loss function, the multi-view loss function, and the regularization penalty to minimize the following overall loss function:

\begin{align}
( \hat{\bm B}^{d,s}, \hat{\bm Z}^s, \hat{\bm \Theta}, \hat{\beta}_0) &= \underset{\bm B^{d,s}, \bm Z^s, \bm \Theta, \beta_0}{\text{min}} \sum_{s=1}^S F(\bm Y^s, \bm Z^s, \bm \Theta, \beta_0) + \sum_{d=1}^D \sum_{s=1}^S F(\bm X^{d,s}, \bm Z^s, \bm B^{d,s}) +  \nonumber\\
& \sum_{d=1}^D\sum_{s=1}^S \mathcal{J}(\bm B^{d,s})
\label{objective}
\end{align}

Although versions of the hierarchical penalty have been used before, our paper is among the first to use this penalty in joint association and prediction studies for data from multiple views to account for common and subgroup-specific variation and to extract subgroup-specific features and/or clinical variables. In Section \ref{sec:alg}, we describe our algorithm for obtaining these estimates. 

% updated
\subsection{Prediction}
\label{sec:pred}
In order to predict an outcome on new data (say $\bm X^{d,s}_{test}$), we first predict the test shared component, $\bm Z^s_{pred}$, and then use this information to predict the outcome. To predict $\bm Z^s_{pred}$, we learn the model defined by (\ref{objective}) on the training data (i.e., $\bm X^{d,s}_{train}$), obtain the learned estimates $\hat{\bm B}^{d,s}$, $\hat{\bm \Theta}$, and $\hat{\beta}_0$. Using these estimates and the testing data $\bm X^{d,s}_{test}$, we solve the problem in \eqref{eq:pred}.

\begin{equation}\label{eq:pred}
    \hat{\bm Z}^s_{pred}= \underset{ \bm Z^s}{\text{min}} \sum_{d=1}^D \sum_{s=1}^S F(\bm X^{d,s}_{test}, \bm Z^s, \hat{\bm B}^{d,s}) = \underset{ \bm Z^s}{\text{min}}\sum_{d=1}^D \sum_{s=1}^S\|\bm X^{d,s}_{test} - \bm Z^s \hat{\bm B}^{{d,s}^T}\|_F^2.
\end{equation}

Without an orthogonality condition on $\bm Z^s$, the solution of this problem has a closed form  given by $\hat{\bm Z}^s_{pred} = \bm X_{cat}^s {\hat{\bm B}_{cat}^s}({\hat{\bm B}_{cat}^{s^T}} {\hat{\bm B}_{cat}^s})^{-1}$
for $s=1,\ldots,S$.  Here, $\bm X_{cat}^s$ is an $n_s \times \{p_1 + \cdots + p_d\}$ matrix that concatenates all $D$ views for subgroup $s$, i.e., $X_{cat}^s=[ \bm X^{1,s}_{test},\cdots,\bm X^{D,s}_{test}]$. Similarly, $\hat{\bm B}_{cat}= [\hat {\bm B}^{1,s},\cdots, \hat{\bm B}^{D,s}]$ and is a $\{p_1 + \cdots + p_D \} \times K$ matrix of variable coefficients. We add a small multiple of the identity matrix before taking the inverse to help with stability, although we note the inverse is of a $K \times K$ matrix, which is computationally inexpensive since $K$ is typically small. With an orthogonality condition on $\bm Z^s$, the above optimization for $\hat{\bm Z}^s_{pred}$ is an orthogonal Procrustes problem \citep{gower2004procrustes}. Let the singular value decomposition of $\bm X_{cat}^s {\hat{\bm B}_{cat}^{s}}$ be $\bm U \bm D \bm V^{T}$. Then $\hat{\bm Z}^s_{pred}  = \bm U \bm V^{T}$. 
Once we have obtained $\hat{\bm Z}^s_{pred}$, we predict a continuous outcome $\hat{\bm Y}^s_{pred}$ as $\hat{\bm Y}^s_{pred} = \hat{\beta}_0 + \hat{\bm Z}_{pred}^s \hat{\bm \Theta}$. 

%-----------------------------------------------------------------------
\section{Algorithm}\label{sec:alg}

% updated
The optimization problem in \eqref{objective} is multi-convex in $\bm B^{d,s}$, $\bm Z^s$ and $\bm \Theta$ each but jointly non-convex. As such, we are not guaranteed convergence to a global minimum. A local optimum can be found by iteratively minimizing over each of the optimization parameters with the rest of the optimization parameters fixed. Overall algorithm convergence is determined by the relative change in the objective function in \eqref{objective} without the penalty terms. The full algorithm is summarized in Algorithm \ref{alg:overview}.

\begin{algorithm}
\caption{Overview of Optimization Algorithm}\label{alg:overview}
\begin{algorithmic}

\State Initialize ${\bm Z^s}^{(0)}$, ${\bm \theta}^{(0)}$, ${\beta_0}^{(0)}$, ${\bm G^d}^{(0)}$, and ${\bm \Xi^{d,s}}^{(0)}$

\For{$t = 1, ..., iter_{max}$}
    %Estimate $Z^s$ using FISTA with backtracking\;
    \For{$s = 1, ..., S$}
        \State ${\bm Z^s}^{(t)} \gets \argmin_{\bm Z^s} F(\bm Y^s, {\bm Z^s}^{(t-1)}, {\beta_0}^{(t-1)}, {\bm \Theta}^{(t-1)}) +  \sum_{d=1}^D F(\bm X^{d,s}, {\bm Z^s}^{(t-1)}, {\bm G^d}^{(t-1)}, {\bm \Xi^{d,s}}^{(t-1)})$ 
        \State Standardize columns of ${\bm Z^s}^{(t)}$ to have mean 0 and variance 1
    \EndFor % End S for Z optimization

    % estimate G^d
    \For{$d = 1, ..., D$}
        \State ${\bm G^d}^{(t)} \gets \argmin_{\bm G^d} \sum_{s=1}^S F(\bm X^{d,s}, {\bm Z^s}^{(t)}, {\bm G^d}^{(t-1)}, {\bm \Xi^{d,s}}^{(t-1)}) + \lambda_g \sum_{l=1}^{p_d} \| {g_l^d}\|_2$ 

        % Estimate Xi^d,s
        \For{$s = 1, ..., S$}
            \State ${\bm \Xi^{d,s}}^{(t)} \gets \argmin_{\bm \Xi^{d,s}}  F(\bm X^{d,s}, {\bm Z^s}^{(t)}, {\bm G^d}^{(t)}, {\bm \Xi^{d,s}}^{(t-1)}) + \lambda_\xi  \sum_{l=1}^{p_d} \| {\xi_l^{d,s}}\|_2$ 
        \EndFor % End S for Xi optimization
    \EndFor % End D for G optimization

    % Estimate theta and beta
    \State $\bm \Theta^{(t)}, \beta_0^{(t)} \gets \argmin_{\bm \Theta, \beta_0} \sum_{s=1}^S F(\bm Y^s, {\bm Z^s}^{(t)}, {\beta_0}^{(t-1)}, {\bm \Theta}^{(t-1)})$ 
    
    % Check converence
    \If{Relative Loss $< \epsilon$}
    %$\frac{|\mathcal{L}^i - \mathcal{L}^{i-1}|}{\mathcal{L}^{i-1}} \leq \epsilon = 10^{-4}$ where $\mathcal{L}^i = \sum_{s=1}^S F( Y^s, {Z^s}^{i}, {\beta_0}^{i}, {\Theta}^{i}) + \sum_{d=1}^D \sum_{s=1}^S F(X^{d,s}, {Z^s}^{i}, {G^d}^{i}, {\Xi^{d,s}}^{i})$
        \State Declare convergence and return estimates
    \ElsIf{$t = iter_{max}$}
        \State Return estimate with warning
    \EndIf
\EndFor % End outer loop iteration
\end{algorithmic}
\end{algorithm}

% updated
\subsection{Optimization Details}

\subsubsection*{Initializations}
The entries of $\bm Z^{s^{(0)}}$ are initialized using random draws from a $U(0.9, 1.1)$ distribution. We initialize the entries of $\bm G^{d^{(0)}}$ for $d = 1, ..., D$,  $\bm \Theta^{(0)}$, and $\beta_0^{(0)}$ with ones. $\bm \Xi^{{d,s}^{(0)}}$ is initialized to minimize $||\bm X_{train}^{d,s} - \bm Z^{s^{(0)}} {\bm \Xi^{{d,s}^{(0)}}}^T||_F^2$, i.e., $\bm \Xi^{{d,s}^{(0)}} = [({\bm Z^{s^{(0)}}}^T \bm Z^{s^{(0)}})^{-1} {\bm Z^{s^{(0)}}}^T \bm X_{train}^{d,s}]^T$.

% updated
\subsubsection*{Estimating \texorpdfstring{$\bm Z^s$}{Zs}} 
After initializations, we first estimate $\hat{\bm Z}^{s^{(t)}}$ by optimizing Equation \eqref{eq:zopt} below using gradient descent with gradients calculated using PyTorch \citep{PyTorch}. 

\begin{equation}\label{eq:zopt}
\hat{\bm Z}^{s^{(t)}} = \underset{\bm Z^s}{\text{min}} \sum_{s=1}^S F(\bm Y^s, \bm Z^s, \bm \Theta^{(t-1)}, \beta_0^{(t-1)}) + \sum_{d=1}^D \sum_{s=1}^S \|\bm X^{d,s} - \bm Z^s \bm B^{{{d,s}^{(t-1)}}^T}\|_F^2
\end{equation}

We use FISTA (fast iterative shrinkage-thresholding algorithm) with backtracking \citep{FISTA} to speed up convergence and select an appropriate step size. FISTA accomplishes the improved complexity by using a linear combination of the previous two iterations when updating optimization parameters rather than just the previous iteration. The convergence criterion is the relative change in \eqref{eq:zopt} evaluated at $\hat{\bm Z}^{s^{(t)}}$ and $\hat{\bm Z}^{s^{(t-1)}}$.

\subsubsection*{Estimating \texorpdfstring{$B^{d,s}$}{B}} 
Estimation of $B^{{d,s}^{(t)}}$ requires first estimating $G^{d^{(t)}}$ for each of the $D$ data views. We fix $\bm \Xi^{{d,s}^{(t-1)}}$ and estimate ${\bm G}^{d^{(t)}}$ by optimizing equation \eqref{eq:gopt} using the Adagrad \citep{Adagrad} optimizer in PyTorch \citep{PyTorch}. We define the convergence criterion as the relative change in \eqref{eq:gopt} evaluated at ${\bm \hat{G}}^{d^{(t)}}$ and ${\bm \hat{G}}^{d^{(t-1)}}$.

\begin{equation}\label{eq:gopt}
    \hat{\bm {G}^d}^{(t)} = \min_{\bm G^d \in R^{p_d \times K}}  \sum_{s=1}^S \|(\bm {X}^{d,s} - \bm Z^{s^{(t)}}({{\bm G^d} \cdot {\bm \Xi^{d,s}}}^{(t)})^T\|^2_F + \lambda_G \gamma_d \sum_{l=1}^{p_d} \| {\bm g_l^d}\|_2
\end{equation}

We then use these updated estimates for $\bm Z^s$ and $\bm G^d$ to estimate ${\bm \Xi^{d,s}}^{(t)}$ by solving equation \eqref{eq:xiopt}. This optimization is performed using the same technique as for $\bm G^d$ with an analogous convergence criterion defined as the relative change in \eqref{eq:xiopt} evaluated at $\bm {\hat{\Xi}}^{{d,s}^{(t)}}$ and $\bm {\hat{\Xi}}^{{d,s}^{(t-1)}}$.

\begin{equation}\label{eq:xiopt}
    {\hat{\bm \Xi}^{{d,s}^{(t+1)}}} = \min_{\bm {\Xi}^{d,s} \in R^{p_d \times K}} \|(\bm {X}^{d,s} - \bm {Z}^{s^{(t)}}(\bm {{G^d}}^{(t)} \cdot {\bm {\Xi}}^{d,s})^T)\|^2_F + \lambda_\xi \gamma_d \sum_{l=1}^{p_d}  \| \bm {\xi}_l^{d,s}\|_2
\end{equation}

Because our implementation uses an automatic differentiation algorithm, the $L_{2,1}$ (or block $l_2/l_1$) penalty does not result in zero coefficients. However, the magnitude of the coefficients in $\hat{\bm B}^{d,s}$ for the noise variables are clearly smaller than the coefficients of the signal variables. We rank and identify important variables based on the magnitude of the $L_2$ norm of the corresponding row in $\hat{\bm B}^{d,s}$.

% updated
\subsubsection*{Estimating  \texorpdfstring{$\Theta$, $\beta_0$}{theta}}
Finally, we update the estimate of $\bm \Theta$ and $\beta_0$ using $\hat{\bm Z^s}^{(t)}$ to optimize equation \eqref{eq:thetaopt}. We use ISTA (iterative shrinkage-thresholding algorithm) with backtracking \citep{FISTA} to select an appropriate step size. The convergence criterion is the relative change in \eqref{eq:thetaopt} evaluated at $\hat{\bm \Theta}^{(t)}, \hat{\beta}_0^{(t)}$ and $\hat{\bm \Theta}^{(t-1)}, \hat{\beta}_0^{(t-1)}$.

\begin{equation}\label{eq:thetaopt}
    \hat{\bm \Theta}^{(t)}, \hat{\beta}_0^{(t)} = \underset{\bm \Theta, \beta_0}{\text{min}} \sum_{s=1}^S F(\bm Y^s, {\bm Z}^{s^{(t)}}, \bm \Theta, \beta_0)
\end{equation}

% addded 
\subsection{Ranking Procedure}\label{ranking}

Because sparsity will not be induced directly due to numerical limitations of the automatic differentiation algorithm, we identify important variables by ranking according to the $L_2$ norm of the corresponding row in $\hat{\bm B}^{d,s}$. The user specifies the number of variables (denote as $N_{top}$) they wish to keep; this value can vary across data views. Algorithm \ref{alg:overview} is run once on the full training data and the $N_{top}$ variables are selected for each view and subgroup based on the estimated $\bm B^{d,s}$. Algorithm \ref{alg:overview} is run a second time including only these selected variables. This `subset' result should be used when applying the prediction procedure in section \ref{sec:pred}.

\subsection{Tuning Parameters}\label{tuning}
The optimization problem depends on tuning parameters $\lambda=(\lambda_G, \lambda_{\xi})$ and the number of latent components ($K$) used to approximate the $\bm X^{d,s}$. First, we use two versions of HIP based on ideas from \cite{bergstra2012random}: (1) HIP (Grid) searches a grid across all points in the hyperparameter space and (2) HIP (Random) searches a random subset of points (parameter combinations) from this grid. The code offers both cross-validation and BIC as methods for selecting $\lambda$. BIC is defined as $N \log\Big(\Big[\sum_{s=1}^S F(\bm Y^s, \hat{\bm Z}^s, \hat{\bm \Theta}, \hat{\beta}_0) + \sum_{d=1}^D \sum_{s=1}^S \| \bm {X}^{d,s} - \hat{\bm Z}^s \hat{\bm B}^{{d,s}^T}\|_F^2\Big]/N \Big) + \log(N) \lambda_B$ where $\lambda_B$ is the sum of the number of non-zero rows across the $\hat{\bm B}^{d,s}$ matrices. Second, we propose an automatic approach to select $K$. Supplementary Tables S1 and S2 present results of how often this approach selects the true value of $K$. Additional simulations found robust results for varying $K$ (Supplementary Figures S1 and S2). Please refer to Section 1 of the Supplementary Information for further discussion of both parameters $\lambda$ and $K$.

%-----------------------------------------------------------------------
\section{Simulations}\label{sec:sim}

% updated
\subsection{Set-up}
\begin{figure}[htbp]
    \centering
    \includegraphics[width=0.5\textwidth]{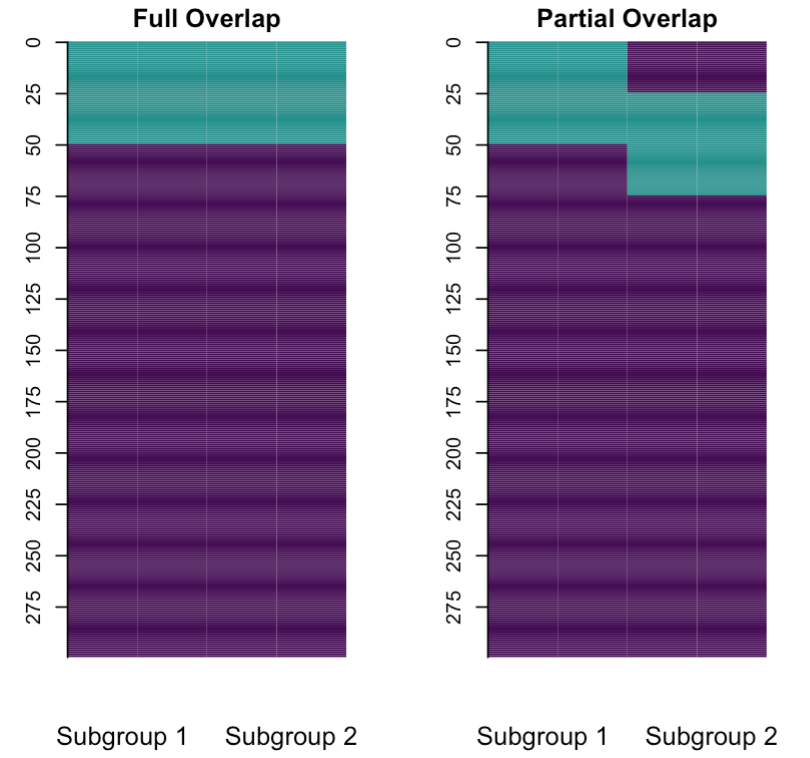}
    \caption{Visual Representation of Variable Overlap Scenarios. In the Full Overlap scenario, the signal variables for each subgroup completely overlapped, i.e., the same variables were important for both subgroups. In the Partial Overlap Scenario, half of the variables important to each subgroup are the same, and the remaining important variables are unique to each subgroup.}\label{fig_settings}
\end{figure}

We performed simulations for a single continuous outcome with two views and two subgroups, i.e., $D=S=2$. There were $n_1 = 250$ subjects in the first subgroup and $n_2 = 260$ subjects in the second. There were two different scenarios to test the ability of the algorithm to perform variable ranking and prediction: Full Overlap and Partial Overlap (Figure \ref{fig_settings}). In the Full Overlap scenario, the signal variables for each subgroup completely overlapped i.e., the first $50$ variables of the $\bm {B}^{d,s}$ matrices were important for both subgroups. We expect competing methods to perform relatively well in this scenario as there is no subgroup heterogeneity. In the Partial Overlap scenario, the first $50$ variables are important for the first subgroup. Of these $50$, the last $25$ are also important to the second subgroup in addition to the $25$ subsequent variables. We expect some deterioration in the ability of the comparison methods to select the appropriate variables due to subgroup heterogeneity.

For each example and scenario, there were three different numbers of variables in the data sets with $p_d$ indicating the number of variables in view $d$. In the P1 setting, $p_1 = 300$ and $p_2 = 350$. In the P2 setting, $p_1 = 1,000$ and $p_2 = 1,500$. Finally, in the P3 setting, $p_1 = 2,000$ and $p_2 = 3,000$. For these simulations, $K$ was fixed to the true value of $2$. We also set $N_{top}$ to the true value of 50 for all simulations.

The data generation process is based on \cite{luo2016canonical}. First, the $\bm {B}^{d,s}$ matrices are generated according to the Full or Partial Overlap scenario. If the entry corresponds to a signal variable, it is drawn from a $U(0.5,1)$ with the sign determined by a draw from a Bernoulli distribution with equal probability; otherwise it is set to 0. We then orthogonalize the columns of each $\bm {B}^{d,s}$. Next, we generate the entries of $\bm Z^s \sim N(\mu = 25.0, \sigma = 3.0)$ and $\bm {E}^{d,s} \sim N(\mu = 0.0, \sigma = 1.0)$. Then the data matrix for subgroup $s$ in view $d$ is generated as $\bm X^{d,s}$ = $\bm Z^s \bm B^{{d,s}^T} + \bm E^{d,s}$. Finally, the outcome is generated as $\bm Y^s = \beta_0 + \bm Z^s \bm \Theta + \bm E^s$ where $\bm E^s \in R^{n_s \times 1}$ contains entries from a standard normal distribution.  The true value of $\bm \Theta = [0.7, 0.2]^T$ and $\beta_0 = 2.0$. 

% updated
\subsection{Comparison Methods}
First, we compare our proposed method (HIP) to canonical variate regression (CVR) \citep{luo2016canonical} as implemented in R package \textit{CVR} \citep{cvrpackage}. This is a joint association and prediction method for multiple views (though existing code only implements two views), but it does not account for subgroup heterogeneity. Thus, we implement the method in two ways: (1) all subgroups are concatenated in each view (Concatenated CVR) and (2) a separate model is fit for each subgroup (Subgroup CVR). Second, we compare our method to the Joint Lasso \citep{jointlasso} as implemented in R package \textit{fuser} \citep{fuser}. The Joint Lasso does not perform integrative analysis but does account for subgroup heterogeneity.  We implement this method on the data stacked over views (Concatenated Joint Lasso), but because Joint Lasso allows for subgroups, we also apply the method on each view separately (Dataset Joint Lasso). Third, we compare our method to the Lasso \citep{lasso} and Elastic Net \citep{enet} as implemented  in R package \textit{glmnet} \citep{glmnet} using both the concatenated and separate subgroup models (Concatenated Lasso/Elastic Net and Subgroup Lasso/Elastic Net respectively); we stack the two views in each case. For the Elastic Net, we fixed $\alpha  = 0.5$. In fitting the models, we allowed any non-fixed tuning parameters to be chosen using the default in the corresponding R package. For Joint Lasso, there is no function for choosing the two tuning parameters in the R package, so we implemented a grid search over 55 parameter combinations. We applied each method to the training data sets and predicted the outcome on the test data sets. We do not compare with the meta lasso because it is only available for binary outcomes and thus not applicable to the motivating COPD data.

% updated
\subsection{Evaluation Measures}
We compare HIP to existing methods in terms of variable selection and prediction. For variable selection, we estimate the true positive rate (TPR), false positive rate (FPR), and F$1$ score. All are constrained to the range $[0,1]$. Note $\text{TPR} = \frac{\text{True Positives}}{\text{True Positives + False Negatives}}$, and $\text{FPR} = \frac{\text{False Positives}}{\text{True Negatives + False Positives}}$. Also, $\text{F1} = \frac{\text{True Positives}}{\text{True Positives} + \frac{1}{2}\text{(False Positives + False Negatives)}}$. Ideally, TPR and F$1$ are 1 and FPR is $0$. 

For HIP, the variables are ranked by the $L_2$ norm of the rows in the estimated $\bm B^{d,s}$ matrices. For each comparison method, the result includes some kind of regression coefficients, so variables with an estimated coefficient that has been shrunk zero are considered not selected and those with non-zero estimated coefficients are considered selected. For prediction, we estimated test mean squared error (MSE); smaller MSEs indicate better performance. We averaged results over our 20 test data sets. 

% updated
\subsection{Results}
\begin{figure}[htbp]
    \centering
    \includegraphics[width=0.85\textwidth]{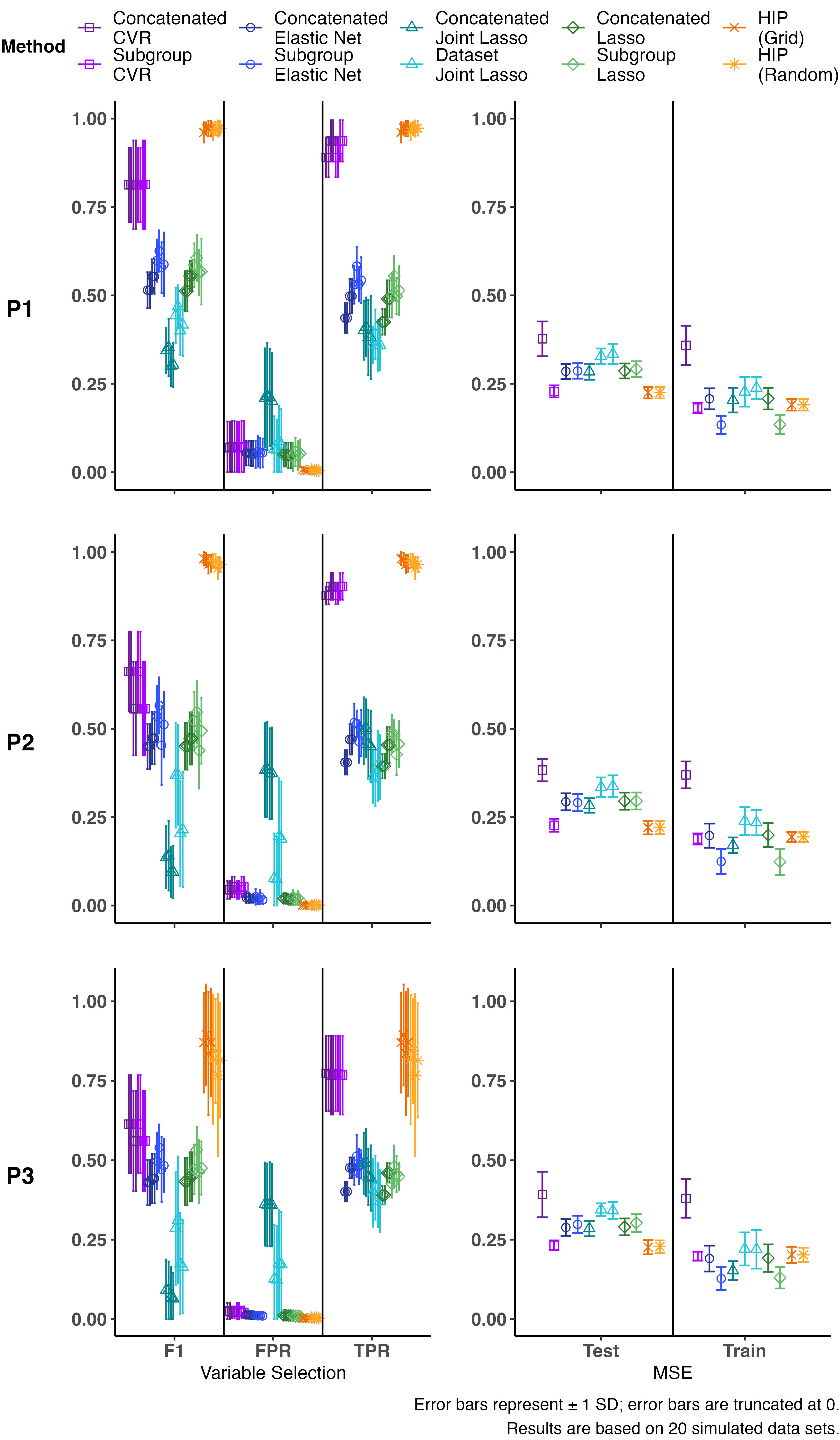}
    \caption{Results for Full Overlap Scenario. The first row corresponds to P1 ($p_1$ = 300, $p_2$ = 350), the second to P2 ($p_1$ = 1000, $p_2$ = 1500), and the third to P3 ($p_1$ = 2000, $p_2$ = 3000). For all settings, $n_1 = 250$ and $n_2 = 260$. The right column is test mean squared error (MSE), so a lower value indicates better performance. All results are based on 20 iterations.}\label{fig_performance_full}
\end{figure}

In the Full Overlap scenario, we compare HIP (Grid) and HIP (Random) and find similar results. This supports using HIP (Random) over HIP (Grid) as it is faster computationally (Supplementary Figure S4). Looking at Figure \ref{fig_performance_full}, we note that HIP has a TPR and F$1$ close to $1$ and FPR close to $0$. CVR is the closest competing method for F$1$ score. For Joint Lasso, the FPR is fairly high, so it is selecting a lot of unimportant variables. Joint Lasso, Lasso, and Elastic Net have lower TPR values suggesting these methods are missing important variables. Overall, the competing methods have worse and more variable TPR, FPR, and F$1$ scores compared to HIP. In terms of prediction, HIP and Subgroup CVR have lower test MSEs than the other methods. Even when subgroups share the same important variables, we see advantages in taking an integrative approach and accounting for heterogeneity. The results are mostly consistent across P1, P2, and P3, although P3 does show some deterioration in performance and increased variability. The Lasso and Elastic net are the fastest in all parameter settings followed by HIP (Random). HIP (Random) shows a larger computational advantage over CVR and Joint Lasso as the number of variables increase (Supplementary Figures S4 and S5). The Partial Overlap scenario results (Supplementary Figure S3) are similar to the Full Overlap results but show a greater advantage for HIP in variable selection performance. 

% Full Overlap Performance

%-----------------------------------------------------------------------
\section{Real Data Analysis}\label{sec:real}

\subsection{Study Goals}
% mostly updated - don't mention all variables in Table 1 explicitly
As mentioned previously, sex disparities exist in COPD susceptibility. In this section, our goal is to use molecular data from the COPDGene Study \citep{regan2011genetic} in combination with clinical data to gain new insights into the molecular architecture of COPD in males and females.  We focus on individuals with COPD (defined as GOLD stage $\ge 1$)  at Year 5 who had proteomics, RNA-sequencing, and AWT data available at Year 5. Of the $N=1376$ individuals with COPD at Year 5 who had complete data, $n_1=782$ were males and $n_2=594$ were females. Table \ref{tab_RDA_dem} gives some characteristics of subjects who had COPD at Year 5. We assessed for sex differences using t-tests for continuous variables and $\chi^2$ tests for categorical variables. Subjects were predominantly non-Hispanic white, but there were no sex differences. There were also not sex differences in age, BMI, systolic blood pressure, percentage of current smokers, or percentage with diabetes. Males and females differed in their mean AWT (p $<0.001$) but did not differ by lung function as measured by mean FEV$_1$\% predicted. Given the available data, and the sex differences in AWT, we will i) identify genes and proteins common and specific to males and females associated with AWT, ii) explore pathways enriched in the proteins and genes identified for males and females, and iii) investigate the effect of these proteins and genes on AWT, adjusting for covariates.

% Table 1 - Demographics
\begin{table}[htbp]
\caption{COPDGene Participant Characteristics. The measurements presented were collected at the Year 5 study visit to align with the collection of proteomic and genomic data collection. }\label{tab_RDA_dem}%
\begin{tabular*}{\textwidth}{@{\extracolsep\fill}lcccccc}
\toprule
\multicolumn{1}{c}{Variable} & \multicolumn{1}{c}{Males} & \multicolumn{1}{c}{Females} & \multicolumn{1}{c}{P-value} \\
  & N = 782 & N = 594 &  \\
\midrule
Age & 68.28 (8.35) & 68.03 (8.36) & 0.581\\
BMI & 28.03 (5.62) & 27.69 (6.59) & 0.317\\
FEV1 \% Predicted & 61.94 (22.97) & 62.91 (22.59) & 0.431\\
BODE Index & 2.45 (2.45) & 2.63 (2.38) & 0.176\\
\% Emphysema & 11.30 (11.86) & 9.39 (11.45) & 0.003\\
Pack Years & 53.05 (26.63) & 47.57 (24.99) & $<$0.001\\
Airway Wall Thickness & 1.17 (0.23) & 1.00 (0.21) & $<$0.001\\
Non-Hispanic White (\%) & 82 & 78 & 0.084\\
Current Smoker (\%) & 66 & 65 & 0.510\\
Diabetes (\%) & 17 & 14 & 0.204\\
\bottomrule
\end{tabular*}
\footnotetext[]{COPD = Chronic Obstructive Pulmonary Disease}
\footnotetext[]{BMI = Body Mass Index}
\footnotetext[]{FEV$_1$ = Forced Expiratory Volume in 1 Second}
\footnotetext[]{BODE = \underline{B}ody mass index, airflow \underline{O}bstruction, \underline{D}yspnea, and \underline{E}xercise capacity}
\end{table}

% updated
\subsection{Applying the proposed and competing methods}\label{rda:apply_methods}
The original data set has 4979 proteins and 19263 RNAseq variables. To reduce dimensionality, we first applied unsupervised filtering to select the 5000 genes and 2000 proteins with the largest standard deviations. To identify ``stable" genes and proteins, i.e., genes and proteins that would consistently be associated with AWT, we generated 50 random splits of the filtered data, stratified by subgroup, such that for each split 75\% of the data was the training data and 25\% was the testing data. Within each split, we performed supervised filtering by regressing AWT on each of the genes and proteins selected by the unsupervised filtering, adjusting for sex, race, and pack years, and retained genes and proteins with potential to explain the variation in AWT (uncorrected p-value $< 0.05$). This means that the variables entering the models could differ for each split of the data. 

To select tuning parameters, we set the range of possible values for $\lambda_G$ and $\lambda_\xi$ in HIP to $(0,2]$ as in the simulations and selected the best model using BIC. Joint Lasso used 10-fold cross-validation over the same grid values used in the simulations. CVR, Lasso, and Elastic Net used 10-fold cross-validation with default settings to select tuning parameters. HIP and CVR both require specification of a rank, i.e., the number of latent components used in the solutions. Our proposed automatic approach (threshold $=0.25$; refer to Section 1.1 of Supplementary Information) on the concatenated data suggested $K=3$. Interestingly, when applied to each $\bm X^{d,s}$ separately, it suggested $K=3$ for the gene data and $K=1$ for the protein data. Supplementary Figure S6 shows the scree plots for both the concatenated and separate $\bm X^{d,s}$. Based on these results and the robustness seen in the sensitivity analyses, we selected $K=3$ components for HIP and CVR. For HIP, we set $N_{top} = 75$ genes and $25$ proteins.

For each split of the data, we applied HIP (Grid), HIP (Random), and the subgroup versions of the competing methods used in the simulations. For Elastic Net and Lasso, we stacked the views and ran separate analyses for males and females. For Joint Lasso, we ran separate analyses for the protein and gene data. For CVR, we ran separate analyses for males and females. We used the selected tuning parameters and testing datasets to predict AWT and estimate test MSEs. We then selected the top 1\% of genes and proteins based on the product of (a) the number of splits in which the variable was included in the $N_{top}$ variables and (b) the proportion of splits in which the variable was included in the $N_{top}$ variables, i.e., the number of splits in which the variable was included in the $N_{top}$ variables divided by the number of splits in which the variable was entered into the model after the supervised filtering; these represent the ``stable" genes and proteins.

\subsection{Results}
\subsubsection{Average mean squared errors, and proteins and genes selected:}
Supplementary Figures S7 and S8 show violin plots of the test MSEs and run times respectively from all 50 splits of the data. The average test MSEs from the splits were slightly lower for CVR and Joint Lasso, but also used many more variables (Supplementary Table S3). HIP (Random) has a computational advantage over CVR and Joint Lasso.

Supplementary Table S4 shows the number of ``stable" common and subgroup-specific genes and proteins identified by each method. We note few overlaps in selected genes and proteins between HIP and existing methods (Supplementary Figure S9). Supplementary Table S5 compares the variables selected by HIP (Random) and HIP (Grid); the selected genes and proteins are very similar, again supporting the use of the random search instead of the grid search. 

Supplementary Tables S6 and S7 list the genes and Supplementary Table S8 lists the proteins identified as ``stable" and important to males and females by HIP (Random) including weights for each protein and gene calculated as the $L_2$ norm of coefficients in $\hat{\bm B}^{d,s}$ across components (i.e., rows) and averaging over the splits where the variable was selected.  

Proteins with large weights include NPLOC4 and SPG21 for males, and SMAP1 and CDKN2D for females. \cite{maxwell_2015_spg21} introduced a novel method called SubmiRine to analyze miRNA and predict miRNA target site variants (miRNA-TSV). When this method was applied to a subset of genomic samples from patients with COPD from the Lung Genome Research Consortium (LGRC; http://www.lung-genomics. org), SPG21 was the top-scoring miRNA-TSV.

The gene with the largest weight was ADIPOR1 for males and BCL2L1 for females. In a study of 60 male COPD patients and 30 male controls, \cite{jaswal_2018_adipor} found adiponectin is associated with inflammation from COPD evidenced by a positive correlation with IL-8 and a negative correlation with FEV$_1$ \%.

\subsubsection{Pathway Enrichment Analysis}
We performed pathway enrichment analysis using Ingenuity Pathway Analysis (IPA) \citep{IPA} to test for overrepresentation of pathways among our lists of ``stable'' proteins and genes for males and females. The top 10 canonical gene pathways (Table \ref{tab_RDA_pathways}) for males and females had some common and some subgroup-specific pathways. The top pathway for males is the iron homeostasis signaling pathway; this is the second ranked pathway for females, and the top pathway for females is heme biosynthesis II. There is strong evidence that disrupted iron homeostasis is associated with the presence and severity of lung disease including COPD \citep{neves_2019_iron, cloonan_2017_iron}. Methylglyoxal degradation I ranks second for males and third for females. \cite{salit_2019_sae} performed gene expression profiling on small airway epithelium samples and also found this pathway to be activated in both male and female smokers.

There was no overlap in the top 10 protein pathways for males and females. The top pathway for males was role of JAK2 in hormone-like cytokine signaling. The top pathway for females was granulocyte adhesion and diapedesis which is associated with regulation of inflammation. \cite{wang_2022_pathway} also found this to be a top pathway involving upregulated genes when comparing patients with COPD and healthy controls. 

% IPA Pathways Table
\begin{sidewaystable}[htbp]
\caption{Top Canonical Pathways from IPA Enrichment Analysis}\label{tab_RDA_pathways}%
\begin{tabular*}{\textwidth}{@{\extracolsep\fill}p{0.25cm}p{0.25cm}p{8cm}p{3.5cm}p{1.25cm}}
\toprule
View & Subgroup &	Canonical Pathway	&	Molecules	&	Unadjusted P-value\\
\midrule
\multirow{20}{*}{\centering Genes}	&	\multirow{10}{*}{\centering Males}	&	Iron homeostasis signaling pathway	&	CDC34,FECH,SLC25A37	&	0.003	\\
	&		&	Methylglyoxal Degradation I	&	HAGH	&	0.006	\\
	&		&	Heme Biosynthesis from Uroporphyrinogen-III I	&	FECH	&	0.008	\\
	&		&	Pentose Phosphate Pathway (Non-oxidative Branch)	&	RPIA	&	0.013	\\
	&		&	Heme Biosynthesis II	&	FECH	&	0.019	\\
	&		&	Pentose Phosphate Pathway	&	RPIA	&	0.023	\\
	&		&	Erythropoietin Signaling Pathway	&	BCL2L1,GATA1	&	0.054	\\
	&		&	ID1 Signaling Pathway	&	BCL2L1,GSPT1	&	0.068	\\
	&		&	Sertoli Cell-Sertoli Cell Junction Signaling	&	SPTB,YBX3	&	0.069	\\
	&		&	Autophagy	&	GABARAPL2,SLC1A5	&	0.076	\\
									
    \cmidrule(lr){2-5}
         
&	\multirow{10}{*}{\centering Females}	&	Heme Biosynthesis II	&	ALAS2,FECH	&	$<$0.001	\\
&		&	Iron homeostasis signaling pathway	&	ALAS2,CDC34,FECH,SLC25A37	&	$<$0.001	\\
&		&	Methylglyoxal Degradation I	&	HAGH	&	0.006	\\
&		&	Heme Biosynthesis from Uroporphyrinogen-III I	&	FECH	&	0.008	\\
&		&	Tetrapyrrole Biosynthesis II	&	ALAS2	&	0.010	\\
&		&	Hypoxia Signaling in the Cardiovascular System	&	CDC34,UBE2H	&	0.011	\\
&		&	Pentose Phosphate Pathway (Non-oxidative Branch)	&	RPIA	&	0.013	\\
&		&	Pentose Phosphate Pathway	&	RPIA	&	0.023	\\
&		&	Erythropoietin Signaling Pathway	&	BCL2L1,GATA1	&	0.054	\\
&		&	ID1 Signaling Pathway	&	BCL2L1,GSPT1	&	0.068	\\

 \midrule 
									
\multirow{20}{*}{\centering Proteins}	&	\multirow{10}{*}{\centering Males}	&	Role of JAK2 in Hormone-like Cytokine Signaling	&	EPO,LEP,PTPN6	&	$<$0.001	\\
	&		&	White Adipose Tissue Browning Pathway	&	BDNF,LEP,NPPB	&	$<$0.001	\\
	&		&	Erythropoietin Signaling Pathway	&	EPO,LEP,PTPN6	&	$<$0.001	\\
	&		&	Serotonin Receptor Signaling	&	ADIPOQ,BDNF,LEP,NPPB	&	$<$0.001	\\
	&		&	AMPK Signaling	&	ADIPOQ,INS,LEP	&	0.001	\\
	&		&	Leptin Signaling in Obesity	&	INS,LEP	&	0.002	\\
	&		&	IL-3 Signaling	&	PPP3R1,PTPN6	&	0.002	\\
	&		&	Maturity Onset Diabetes of Young (MODY) Signaling	&	ADIPOQ,INS	&	0.002	\\
	&		&	Thyroid Cancer Signaling	&	BDNF,INS	&	0.002	\\
	&		&	ABRA Signaling Pathway	&	NPPB,PPP3R1	&	0.003	\\

   \cmidrule(lr){2-5}
									
	&	\multirow{10}{*}{\centering Females}	&	Granulocyte Adhesion and Diapedesis	&	PF4,PPBP,TNFRSF1A	&	$<$0.001	\\
	&		&	Agranulocyte Adhesion and Diapedesis	&	PF4,PPBP,TNFRSF1A	&	$<$0.001	\\
	&		&	Wound Healing Signaling Pathway	&	EGF,PF4,TNFRSF1A	&	$<$0.001	\\
	&		&	Huntington's Disease Signaling	&	BDNF,CPLX2,EGF	&	0.001	\\
	&		&	Pathogen Induced Cytokine Storm Signaling Pathway	&	PF4,PPBP,TNFRSF1A	&	0.002	\\
	&		&	Glioma Signaling	&	CDKN2D,EGF	&	0.004	\\
	&		&	Type II Diabetes Mellitus Signaling	&	ADIPOQ,TNFRSF1A	&	0.005	\\
	&		&	Axonal Guidance Signaling	&	BDNF,EGF,PAPPA	&	0.005	\\
	&		&	Tumor Microenvironment Pathway	&	EGF,TNFRSF1A	&	0.007	\\
	&		&	Regulation Of The Epithelial Mesenchymal Transition By Growth Factors Pathway	&	EGF,TNFRSF1A	&	0.008	\\
\bottomrule
\end{tabular*}
\end{sidewaystable}

\subsubsection{Effect of common and sex-specific genes and proteins on AWT}
Finally, we created common and sex-specific protein and gene scores from the ``stable'' proteins and genes selected by HIP (Random) and assessed whether these scores improved the prediction of AWT beyond some established COPD risk factors. We created the common protein score for subject $i$ as CommonProtScore$_{i}=\sum_{j=1}^{\# common~proteins}w_jx_{ij}^1$ where $x_{ij}^1$ is subject $i$'s protein expression value for the $j$th common protein (i.e., the $ij$th entry for  the protein data, $\bm X^1$), and $w_j$ is the weight for protein $j$. Each protein weight, $w_j$, was obtained via bootstrap. Specifically, we obtained 200 bootstrap datasets, and for each bootstrap dataset, we obtained regression coefficients and standard errors from univariate regression models of AWT and each of the common proteins identified. This resulted in 200 regression coefficients and standard errors which we combined using a weighted mean. The subgroup-specific scores were also obtained in a similar fashion. The scores were standardized to have mean 0 and variance 1 in each subgroup since different variables were identified for males and females.  

Once the scores were created, we fit several multiple linear regression models on the full data: (1) Established Risk Factors (ERF) Model, (2) ERF + Common Protein Score, (3) ERF + Common Gene Score, (4) ERF + Common Protein and Gene Scores, (5) ERF + Subgroup Protein Score, (6) ERF + Subgroup Gene Score, (7) ERF + Subgroup Protein and Gene Scores. Table \ref{tab_RDA_reg} shows the coefficient estimates with confidence intervals and p-values. We observe both the common and subgroup-specific protein scores are statistically significant, but neither the common nor subgroup-specific gene scores were. This could be due to including too few genes in the scores or because there was a large overlap between the genes selected for males and females. The ``stable'' proteins we identified to be common and specific to males and females could potentially be explored to further our understanding of sex differences in COPD mechanisms. 

% Regression table on full data
\renewcommand{\arraystretch}{0.5}
\begin{table}[htbp]
\caption{Comparison of Regression Model Estimates. Models were fit on all participants in the COPD Data set and adjusted for age, sex, race, BMI, smoking status, percent emphysema, and scanner make ($N=1374$). Scores were developed using the ``stable" proteins and genes selected by HIP (Random).}\label{tab_RDA_reg}%
\begin{tabular*}{\textwidth}{@{\extracolsep\fill}cccccc}
\toprule
Variable & Estimate & 95\% CI & P-value & $R^2$ & Adjusted $R^2$\\
\midrule
\multicolumn{1}{l}{\textbf{ERF}} &  &  &  & 0.152 & 0.147\\
Intercept & -1.180 & -1.872, -0.489 & 0.001 &  & \\
Age & 0.021 & -0.035,  0.077 & 0.461 &  & \\
Sex (Female) & 0.006 & -0.093,  0.105 & 0.901 &  & \\
Race (African American) & -0.070 & -0.202,  0.062 & 0.297 &  & \\
BMI & 0.352 & 0.299,  0.406 & $<$0.001 &  & \\
Former Smoker & 0.947 & 0.257,  1.637 & 0.007 &  & \\
Current Smoker & 1.429 & 0.735,  2.123 & $<$0.001 &  & \\
\% Emphysema & 0.023 & -0.032,  0.079 & 0.414 &  & \\
Scanner - Philips & 0.379 & 0.097,  0.661 & 0.009 &  & \\
Scanner - Siemens & 0.130 & 0.025,  0.235 & 0.015 &  & \\
\multicolumn{1}{l}{\textbf{ERF + Common Protein Score}} &  &  &  & 0.157 & 0.151\\
Intercept & -1.138 & -1.828, -0.447 & 0.001 &  & \\
Age & 0.032 & -0.024,  0.089 & 0.266 &  & \\
Sex (Female) & 0.007 & -0.092,  0.105 & 0.897 &  & \\
Race (African American) & -0.049 & -0.181,  0.084 & 0.469 &  & \\
BMI & 0.327 & 0.271,  0.383 & $<$0.001 &  & \\
Former Smoker & 0.911 & 0.223,  1.600 & 0.010 &  & \\
Current Smoker & 1.390 & 0.697,  2.083 & $<$0.001 &  & \\
\% Emphysema & 0.028 & -0.028,  0.083 & 0.328 &  & \\
Scanner - Philips & 0.404 & 0.123,  0.686 & 0.005 &  & \\
Scanner - Siemens & 0.110 & 0.005,  0.216 & 0.041 &  & \\
Common Protein Score & 0.076 & 0.023,  0.130 & 0.005 &  & \\
\multicolumn{1}{l}{\textbf{ERF + Common Gene Score}} &  &  &  & 0.153 & 0.147\\
Intercept & -1.169 & -1.861, -0.477 & 0.001 &  & \\
Age & 0.021 & -0.035,  0.077 & 0.462 &  & \\
Sex (Female) & 0.007 & -0.092,  0.106 & 0.893 &  & \\
Race (African American) & -0.083 & -0.216,  0.050 & 0.221 &  & \\
BMI & 0.343 & 0.288,  0.398 & $<$0.001 &  & \\
Former Smoker & 0.932 & 0.242,  1.623 & 0.008 &  & \\
Current Smoker & 1.423 & 0.729,  2.117 & $<$0.001 &  & \\
\% Emphysema & 0.023 & -0.033,  0.079 & 0.421 &  & \\
Scanner - Philips & 0.387 & 0.105,  0.669 & 0.007 &  & \\
Scanner - Siemens & 0.134 & 0.029,  0.239 & 0.013 &  & \\
Common Gene Score & 0.035 & -0.017,  0.086 & 0.185 &  & \\
\multicolumn{1}{l}{\textbf{ERF + Common Scores}} &  &  &  & 0.158 & 0.151\\
Intercept & -1.128 & -1.819, -0.437 & 0.001 &  & \\
Age & 0.032 & -0.025,  0.088 & 0.270 &  & \\
Sex (Female) & 0.007 & -0.092,  0.106 & 0.890 &  & \\
Race (African American) & -0.061 & -0.195,  0.073 & 0.373 &  & \\
BMI & 0.320 & 0.262,  0.377 & $<$0.001 &  & \\
Former Smoker & 0.899 & 0.210,  1.588 & 0.011 &  & \\
Current Smoker & 1.385 & 0.692,  2.078 & $<$0.001 &  & \\
\% Emphysema & 0.027 & -0.028,  0.083 & 0.335 &  & \\
Scanner - Philips & 0.411 & 0.129,  0.693 & 0.004 &  & \\
Scanner - Siemens & 0.114 & 0.008,  0.220 & 0.035 &  & \\
Common Protein Score & 0.074 & 0.021,  0.128 & 0.006 &  & \\
Common Gene Score & 0.031 & -0.020,  0.083 & 0.237 &  & \\
\multicolumn{1}{l}{\textbf{ERF + Subgroup Protein Score}} &  &  &  & 0.165 & 0.159\\
Intercept & -1.071 & -1.760, -0.383 & 0.002 &  & \\
Age & 0.014 & -0.041,  0.070 & 0.614 &  & \\
Sex (Female) & 0.007 & -0.091,  0.105 & 0.891 &  & \\
Race (African American) & -0.035 & -0.167,  0.097 & 0.605 &  & \\
BMI & 0.312 & 0.256,  0.368 & $<$0.001 &  & \\
Former Smoker & 0.860 & 0.174,  1.546 & 0.014 &  & \\
Current Smoker & 1.335 & 0.645,  2.025 & $<$0.001 &  & \\
\% Emphysema & 0.030 & -0.025,  0.086 & 0.284 &  & \\
Scanner - Philips & 0.418 & 0.137,  0.698 & 0.004 &  & \\
Scanner - Siemens & 0.079 & -0.027,  0.185 & 0.146 &  & \\
Subgroup Protein Score & 0.126 & 0.072,  0.179 & $<$0.001 &  & \\
\multicolumn{1}{l}{\textbf{ERF + Subgroup Gene Score}} &  &  &  & 0.153 & 0.147\\
Intercept & -1.169 & -1.861, -0.477 & 0.001 &  & \\
Age & 0.021 & -0.035,  0.077 & 0.459 &  & \\
Sex (Female) & 0.007 & -0.092,  0.106 & 0.893 &  & \\
Race (African American) & -0.083 & -0.217,  0.050 & 0.220 &  & \\
BMI & 0.343 & 0.288,  0.398 & $<$0.001 &  & \\
Former Smoker & 0.932 & 0.242,  1.622 & 0.008 &  & \\
Current Smoker & 1.423 & 0.729,  2.117 & $<$0.001 &  & \\
\% Emphysema & 0.023 & -0.033,  0.079 & 0.421 &  & \\
Scanner - Philips & 0.387 & 0.105,  0.669 & 0.007 &  & \\
Scanner - Siemens & 0.134 & 0.029,  0.239 & 0.013 &  & \\
Subgroup Gene Score & 0.035 & -0.016,  0.087 & 0.180 &  & \\
\multicolumn{1}{l}{\textbf{ERF + Subgroup Scores}} &  &  &  & 0.166 & 0.159\\
Intercept & -1.064 & -1.752, -0.375 & 0.003 &  & \\
Age & 0.015 & -0.041,  0.070 & 0.610 &  & \\
Sex (Female) & 0.007 & -0.091,  0.106 & 0.885 &  & \\
Race (African American) & -0.045 & -0.179,  0.088 & 0.504 &  & \\
BMI & 0.306 & 0.249,  0.363 & $<$0.001 &  & \\
Former Smoker & 0.850 & 0.164,  1.536 & 0.015 &  & \\
Current Smoker & 1.332 & 0.642,  2.022 & $<$0.001 &  & \\
\% Emphysema & 0.030 & -0.025,  0.085 & 0.290 &  & \\
Scanner - Philips & 0.423 & 0.142,  0.704 & 0.003 &  & \\
Scanner - Siemens & 0.083 & -0.024,  0.189 & 0.129 &  & \\
Subgroup Protein Score & 0.124 & 0.070,  0.177 & $<$0.001 &  & \\
Subtype Gene Score & 0.027 & -0.024,  0.078 & 0.304 &  & \\
\bottomrule
\end{tabular*}
\footnotetext[]{ERF = Established Risk Factors (Age, Sex, Race, BMI, and Smoking Status)}
\footnotetext[]{BMI = Body Mass Index}
\end{table}

% Protein and Gene Scores on 100 splits
%\input{Tables/Tab_RDA_scores}

%-----------------------------------------------------------------------
\section{Conclusion}\label{sec:conc}

We have tackled the problem of accounting for subgroup heterogeneity in an integrative analysis framework. Motivated by the COPDGene study and a scientific need to understand sex differences in COPD, we developed appropriate statistical methods that leverage the strengths of multi-view data, account for subgroup heterogeneity, incorporate clinical covariates, and combine the association step with a clinical outcome step to guide the selection of clinically meaningful molecular signatures. Through the use of a hierarchical penalty, we identify omics signatures that are common and subgroup-specific and can predict a clinical outcome. HIP showed comparable to substantially improved prediction and variable selection performance in simulation settings when compared to existing methods.

When we applied HIP to genomic and proteomic data from COPDGene, we identified protein and gene biomarkers and pathways common and specific to males and females. When the proteins and genes were developed into scores, the common and subgroup-specific protein scores were statistically significant predictors of airway wall thickness (AWT) even when including established risk factors of COPD. 
These findings suggest the proteins and genes identified to be common and specific to males and females could be explored to further our understanding of sex differences in COPD mechanisms.

Recently, \cite{Yun_2022_AWTcompare} also explored gene signatures related to AWT and found that interferon stimulated genes were associated with AWT. We did not find these same genes in our analysis, but there were several differences in the analyses that could explain the differing results: (1) the subset of COPDGene participants in the two analyses were different as we only included participants with COPD while \cite{Yun_2022_AWTcompare} included participants with and without COPD, (2) \cite{Yun_2022_AWTcompare} looked for associations between individual genes and AWT while adjusting for covariates whereas we selected genes based on rankings from our model that included several genes at once, (3) we considered both gene and protein data (which also impacted which participants we could include) whereas \cite{Yun_2022_AWTcompare} only considered genes, and (4) we use IPA \citep{IPA} to find pathways whereas \cite{Yun_2022_AWTcompare} used MSigDB (\url{https://www.gsea-msigdb.org/gsea/msigdb}). 

HIP has some limitations warranting further research. First, the number of variables to be kept for the subset model refit has to be specified. In simulations where this value is known, performance is very good, but the truth will not be known in applied settings. Users could look at plots of the weights from the $\hat{\bm B}^{d,s}$ to see how many variables seem to have large weights. We also found that if there were some splits where the train MSEs were very small but test MSEs very large, i.e., evidence of overfitting that more variables needed to be retained. Second, the tuning range for $\lambda_\xi$ and $\lambda_g$ is not determined by the data, so the tuning range may need to be adjusted to attain optimal sparsity. This can be done with an optional parameter in the code. Additionally, the number of  components, $K$, needs to be specified. Although the truth can never be known, we provide an automatic method to select $K$ and discuss other options in the supplemental material. Future research should explore the possibility that $K$ may differ by data view. Finally, HIP is limited to cross-sectional data, but future work could extend it to accommodate longitudinal data to determine whether trends in some outcome vary by subgroup. Despite these limitations, HIP advances statistical methods for joint association and prediction of multi-view data, and the encouraging simulation and real data findings motivate further applications.

%-----------------------------------------------------------------------

\backmatter

\section*{Declarations}
% https://bmcbioinformatics.biomedcentral.com/submission-guidelines/preparing-your-manuscript/research-article

\bmhead{Ethics approval and consent to participate}
This research uses previously collected, de-identified data from the COPDGene Study \citep{regan2011genetic}, a multi-center study with 21 clinical sites each with local IRB approval (NCT00608764).

\bmhead{Consent for publication}
Not applicable

\bmhead{Availability of data and materials}
Access to the clinical and genomic data can be requested through dbGaP (IDs: phs000951.v4.p4 and phs000179.v6.p2). The proteomic data can be requested from the COPDGene Study Group (\url{http://www.copdgene.org/}). 

The Python source code for implementing the methods and generating simulated data along with README files will be available on GitHub at \url{https://github.com/lasandrall/HIP}.

\bmhead{Competing interests}
The authors declare that they have no competing interests.

\bmhead{Funding}
This work was supported by National Center For Advancing Translational Science [5KL2TR002492-04] and National Institute Of General Medical Sciences [1R35GM142695-01].

\bmhead{Authors' contributions}
SES and QL conceived of the idea. SES, JB, and LE developed the methods. JB and SES developed code to implement the methods. JB conducted simulations and real data analyses. JB and CW interpreted results from the real data analyses. JB and SES wrote a first draft of the paper. All authors read and edited the final manuscript.

\bmhead{Disclaimer} The views expressed in this article are those of the authors and do not reflect the views of the United States Government, the Department of Veterans Affairs, the funders, the sponsors, or any of the authors' affiliated academic institutions.

\bmhead{Acknowledgements}
This work was supported by NHLBI U01 HL089897 and U01 HL089856. The COPDGene study (NCT00608764) is also supported by the COPD Foundation through contributions made to an Industry Advisory Committee that has included AstraZeneca, Bayer Pharmaceuticals, Boehringer-Ingelheim, Genentech, GlaxoSmithKline, Novartis, Pfizer, and Sunovion.

\bibliography{References}

% If additional material is provided, please list the following information in a separate section of the manuscript text:
%  - File name (e.g. Additional file 1)
%  - File format including the correct file extension for example .pdf, .xls, .txt, .pptx (including name and a URL of an appropriate viewer if format is unusual)
%  - Title of data
%  - Description of data
%Additional files should be named "Additional file 1" and so on and should be referenced explicitly by file name within the body of the article, e.g. 'An additional movie file shows this in more detail [see Additional file 1]'.

\end{document}

% \title{Supplementary Information for HIP: a method for high-dimensional multi-view data integration and prediction accounting for subgroup heterogeneity}
\title{Supplementary Information for Accounting for data heterogeneity in integrative analysis and prediction methods: An application to Chronic Obstructive Pulmonary Disease}

\date{}
\bigskip
\maketitle

% Relabels as Table/Figure SX for supplement
\renewcommand{\thetable}{S\arabic{table}}
\renewcommand{\thefigure}{S\arabic{figure}}

\section{Additional Methodological Details}\label{appendix_method}

% updated
\subsection{K Selection Approaches}\label{appendix:k_approach}

Selecting the number of components ($K$) is an area of ongoing research, e.g., \cite{bai2002rank} suggest multiple ways to select this value, but all of these methods require running the model for several $K$ values which is computationally infeasible in this framework. Instead, we propose an automatic and computationally efficient method for selecting the number of components, $K$. First, we combine all the views and subgroups into an $N \times \{p_1 + \cdots + p_D\}$ matrix and use the singular values of the correlation matrix to calculate the eigenvalues of these concatenated data. These eigenvalues correspond to the amount of variation explained by the corresponding components. We start with the first component, looking at how much additional variation is explained by adding a second component (how much smaller the second eigenvalue is compared to the first). If the amount of additional variation explained is above some threshold, then we retain this component and repeat the process with the next components until adding an additional component does not explain much additional variation. The amount of additional variation explained will decrease monotonically, which can be seen in a scree plot of the eigenvalues. As such, another option to select $K$ is to make a manual selection based on a scree plot of the concatenated data eigenvalues. We can also either apply the automatic selection or generate scree plots separately for the individual data views for each subgroup, i.e., for each $\bm X^{d,s}$. If the automatic selection is applied to the separate $\bm X^{d,s}$, we take the mode of the returned $K$ values. If there is more than one mode, we recommend selecting the larger of the two modes.

\begin{algorithm}
\caption{Algorithm for Proposed Automatic $K$ Selection}\label{alg:kselect}
\begin{algorithmic}
\State $EV$ is a vector of eigenvalues whose first element is indexed as `0'
\For{$j = 1, ..., p_d - 1$}
    \State{val $\gets EV_j$ / $\sum_{l = 0}^{j} EV_l$}
    \If{val $<$ threshold}
        \State\Return{j} 
    \EndIf
\EndFor 
\end{algorithmic}
\end{algorithm}

%  updated, but still need to summarize results
\subsection{K Selection Simulations and Sensitivity Results}\label{appendix:k_results}

To test the ability of our proposed automatic approach to select the true value of $K$, we applied this approach to the 20 generated data sets for both the Full and Partial Overlap scenarios for the P1, P2, and P3 variable dimensions to both the concatenated data and separate $\bm X^{d,s}$. All data were simulated with a true value of $K = 2$. Tables \ref{tab_Kselect_Full} and \ref{tab_Kselect_Partial} report the distribution of $K$ values selected for the Full and Partial overlap scenarios respectively for threshold values of 0.10, 0.15, 0.20, and 0.25. For the separate data approach, we used the mode as the selected $K$ value. If there was more than one mode, we selected the largest value.

In the Full Overlap scenario using the concatenated data, a threshold of 0.20 or 0.25 for the P1 and P2 dimensions primarily selected 4 components, and a threshold of 0.25 for the P3 dimension primarily selected 4 components. Using the separate data, a threshold of 0.20 or 0.25 primarily selected 3 components for the P1 dimension, and a threshold of 0.25 selected 4 components for the P2 and P3 dimensions. For both the concatenated and separate data, the number of components selected tended to increase as the variable dimensions increased. 

In the Partial Overlap scenario using the concatenated data, a threshold of 0.15, 0.20 or 0.25 for the P1 dimension selected 4 components, and a threshold of 0.25 selected 4 components for the P2 and P3 dimension. Using the separate data, a threshold of 0.20 or 0.25 primarily selected 3 components for the P1 dimension, and a threshold of 0.25 selected 4 components for the P2 and P3 dimensions. Again, for both the concatenated and separate data, the number of components selected tended to increase as the variable dimensions increased.  

%% Full Overlap Table
\begin{table}[htbp]
\caption{K Selection Simulation Results, Full Overlap Scenario. 
We calculated eigenvalues of the correlation matrix of either the concatenated $\bm X$ matrix or each $\bm X^{d,s}$ matrix separately. We look at the percent of additional variation that is explained from one component to the next adding components until the additional proportion of variance explained is less than the given threshold. The `K Selected' columns are the proportion of times that value of K was selected for the given data approach, setting, and threshold. Results are based on 20 simulated data sets. Variable dimensions are P1 = [300, 350], P2 = [1000, 1500], and P3 = [2000, 3000]. Of these variables, the same 50 variables are important to each subgroup, and there are no variables unique to each subgroup.}\label{tab_Kselect_Full}%
\footnotesize
\begin{tabular}{@{}cccccccccccc@{}}
\toprule
\multicolumn{3}{c}{ } & \multicolumn{9}{c}{K Selected} \\
\cmidrule(l{3pt}r{3pt}){4-12}
Data & Setting & Threshold & 2 & 3 & 4 & 5 & 6 & 7 & 8 & 9 & 10\\
\midrule
 &  & 0.10 & 0.00 & 0.00 & 0.00 & 0.00 & 1.00 & 0.00 & 0.00 & 0.00 & 0.00\\

 &  & 0.15 & 0.00 & 0.00 & 1.00 & 0.00 & 0.00 & 0.00 & 0.00 & 0.00 & 0.00\\

 &  & 0.20 & 0.00 & 0.00 & 1.00 & 0.00 & 0.00 & 0.00 & 0.00 & 0.00 & 0.00\\

 & \multirow[t]{-4}{*}{\centering\arraybackslash Full P1} & 0.25 & 0.00 & 0.20 & 0.80 & 0.00 & 0.00 & 0.00 & 0.00 & 0.00 & 0.00\\
\cmidrule{2-12}
 &  & 0.10 & 0.00 & 0.00 & 0.00 & 0.00 & 0.00 & 0.00 & 0.00 & 1.00 & 0.00\\

 &  & 0.15 & 0.00 & 0.00 & 0.00 & 0.00 & 1.00 & 0.00 & 0.00 & 0.00 & 0.00\\

 &  & 0.20 & 0.00 & 0.00 & 0.90 & 0.10 & 0.00 & 0.00 & 0.00 & 0.00 & 0.00\\

 & \multirow[t]{-4}{*}{\centering\arraybackslash Full P2} & 0.25 & 0.00 & 0.00 & 1.00 & 0.00 & 0.00 & 0.00 & 0.00 & 0.00 & 0.00\\
\cmidrule{2-12}
 &  & 0.10 & 0.00 & 0.00 & 0.00 & 0.00 & 0.00 & 0.00 & 0.00 & 0.00 & 1.00\\

 &  & 0.15 & 0.00 & 0.00 & 0.00 & 0.00 & 0.00 & 1.00 & 0.00 & 0.00 & 0.00\\

 &  & 0.20 & 0.00 & 0.00 & 0.00 & 1.00 & 0.00 & 0.00 & 0.00 & 0.00 & 0.00\\

\multirow[t]{-12}{*}{\centering\arraybackslash Concatenated} & \multirow[t]{-4}{*}{\centering\arraybackslash Full P3} & 0.25 & 0.00 & 0.00 & 1.00 & 0.00 & 0.00 & 0.00 & 0.00 & 0.00 & 0.00\\
\cmidrule{1-12}
 &  & 0.10 & 0.00 & 0.00 & 0.00 & 0.00 & 0.00 & 0.00 & 1.00 & 0.00 & 0.00\\

 &  & 0.15 & 0.00 & 0.00 & 0.00 & 1.00 & 0.00 & 0.00 & 0.00 & 0.00 & 0.00\\

 &  & 0.20 & 0.00 & 0.80 & 0.20 & 0.00 & 0.00 & 0.00 & 0.00 & 0.00 & 0.00\\

 & \multirow[t]{-4}{*}{\centering\arraybackslash Full P1} & 0.25 & 0.15 & 0.85 & 0.00 & 0.00 & 0.00 & 0.00 & 0.00 & 0.00 & 0.00\\
\cmidrule{2-12}
 &  & 0.10 & 0.00 & 0.00 & 0.00 & 0.00 & 0.00 & 0.00 & 0.00 & 0.20 & 0.80\\

 &  & 0.15 & 0.00 & 0.00 & 0.00 & 0.00 & 0.70 & 0.30 & 0.00 & 0.00 & 0.00\\

 &  & 0.20 & 0.00 & 0.00 & 0.00 & 1.00 & 0.00 & 0.00 & 0.00 & 0.00 & 0.00\\

 & \multirow[t]{-4}{*}{\centering\arraybackslash Full P2} & 0.25 & 0.00 & 0.00 & 1.00 & 0.00 & 0.00 & 0.00 & 0.00 & 0.00 & 0.00\\
\cmidrule{2-12}
 &  & 0.10 & 0.00 & 0.00 & 0.00 & 0.00 & 0.00 & 0.00 & 0.00 & 0.00 & 1.00\\

 &  & 0.15 & 0.00 & 0.00 & 0.00 & 0.00 & 0.00 & 1.00 & 0.00 & 0.00 & 0.00\\

 &  & 0.20 & 0.00 & 0.00 & 0.00 & 1.00 & 0.00 & 0.00 & 0.00 & 0.00 & 0.00\\

\multirow[t]{-12}{*}{\centering\arraybackslash Separate} & \multirow[t]{-4}{*}{\centering\arraybackslash Full P3} & 0.25 & 0.00 & 0.00 & 1.00 & 0.00 & 0.00 & 0.00 & 0.00 & 0.00 & 0.00\\
\bottomrule
\end{tabular}
\end{table}

%% Partial Overlap Table
\begin{table}[htbp]
\caption{K Selection Simulation Results, Partial Overlap Scenario. 
We calculated eigenvalues of the correlation matrix of either the concatenated $\bm X$ matrix or each $\bm X^{d,s}$ matrix separately. We look at the percent of additional variation that is explained from one component to the next adding components until the additional proportion of variance explained is less than the given threshold. The `K Selected' columns are the proportion of times that value of K was selected for the given data approach, setting, and threshold. Results are based on 20 simulated data sets. Variable dimensions are P1 = [300, 350], P2 = [1000, 1500], and P3 = [2000, 3000]. Of these variables, 25 are important to each subgroup, and there are 25 variables that are unique to each subgroup.}\label{tab_Kselect_Partial}%
\footnotesize
\begin{tabular}{@{}cccccccccccc@{}}
\toprule
\multicolumn{3}{c}{ } & \multicolumn{9}{c}{K Selected} \\
\cmidrule(l{3pt}r{3pt}){4-12}
Data & Setting & Threshold & 2 & 3 & 4 & 5 & 6 & 7 & 8 & 9 & 10\\
\midrule
 &  & 0.10 & 0.00 & 0.00 & 0.00 & 0.00 & 1.00 & 0.00 & 0.00 & 0.00 & 0.00\\

 &  & 0.15 & 0.00 & 0.00 & 1.00 & 0.00 & 0.00 & 0.00 & 0.00 & 0.00 & 0.00\\

 &  & 0.20 & 0.00 & 0.00 & 1.00 & 0.00 & 0.00 & 0.00 & 0.00 & 0.00 & 0.00\\

 & \multirow[t]{-4}{*}{\centering\arraybackslash Partial P1} & 0.25 & 0.00 & 0.00 & 1.00 & 0.00 & 0.00 & 0.00 & 0.00 & 0.00 & 0.00\\
\cmidrule{2-12}
 &  & 0.10 & 0.00 & 0.00 & 0.00 & 0.00 & 0.00 & 0.00 & 0.00 & 1.00 & 0.00\\

 &  & 0.15 & 0.00 & 0.00 & 0.00 & 0.00 & 1.00 & 0.00 & 0.00 & 0.00 & 0.00\\

 &  & 0.20 & 0.00 & 0.00 & 0.75 & 0.25 & 0.00 & 0.00 & 0.00 & 0.00 & 0.00\\

 & \multirow[t]{-4}{*}{\centering\arraybackslash Partial P2} & 0.25 & 0.00 & 0.00 & 1.00 & 0.00 & 0.00 & 0.00 & 0.00 & 0.00 & 0.00\\
\cmidrule{2-12}
 &  & 0.10 & 0.00 & 0.00 & 0.00 & 0.00 & 0.00 & 0.00 & 0.00 & 0.00 & 1.00\\

 &  & 0.15 & 0.00 & 0.00 & 0.00 & 0.00 & 0.00 & 1.00 & 0.00 & 0.00 & 0.00\\

 &  & 0.20 & 0.00 & 0.00 & 0.00 & 1.00 & 0.00 & 0.00 & 0.00 & 0.00 & 0.00\\

\multirow[t]{-12}{*}{\centering\arraybackslash Concatenated} & \multirow[t]{-4}{*}{\centering\arraybackslash Partial P3} & 0.25 & 0.00 & 0.00 & 1.00 & 0.00 & 0.00 & 0.00 & 0.00 & 0.00 & 0.00\\
\cmidrule{1-12}
 &  & 0.10 & 0.00 & 0.00 & 0.00 & 0.00 & 0.00 & 0.00 & 1.00 & 0.00 & 0.00\\

 &  & 0.15 & 0.00 & 0.00 & 0.00 & 1.00 & 0.00 & 0.00 & 0.00 & 0.00 & 0.00\\

 &  & 0.20 & 0.00 & 0.80 & 0.20 & 0.00 & 0.00 & 0.00 & 0.00 & 0.00 & 0.00\\

 & \multirow[t]{-4}{*}{\centering\arraybackslash Partial P1} & 0.25 & 0.10 & 0.90 & 0.00 & 0.00 & 0.00 & 0.00 & 0.00 & 0.00 & 0.00\\
\cmidrule{2-12}
 &  & 0.10 & 0.00 & 0.00 & 0.00 & 0.00 & 0.00 & 0.00 & 0.00 & 0.10 & 0.90\\

 &  & 0.15 & 0.00 & 0.00 & 0.00 & 0.00 & 0.75 & 0.25 & 0.00 & 0.00 & 0.00\\

 &  & 0.20 & 0.00 & 0.00 & 0.00 & 1.00 & 0.00 & 0.00 & 0.00 & 0.00 & 0.00\\

 & \multirow[t]{-4}{*}{\centering\arraybackslash Partial P2} & 0.25 & 0.00 & 0.00 & 1.00 & 0.00 & 0.00 & 0.00 & 0.00 & 0.00 & 0.00\\
\cmidrule{2-12}
 &  & 0.10 & 0.00 & 0.00 & 0.00 & 0.00 & 0.00 & 0.00 & 0.00 & 0.00 & 1.00\\

 &  & 0.15 & 0.00 & 0.00 & 0.00 & 0.00 & 0.00 & 1.00 & 0.00 & 0.00 & 0.00\\

 &  & 0.20 & 0.00 & 0.00 & 0.00 & 1.00 & 0.00 & 0.00 & 0.00 & 0.00 & 0.00\\

\multirow[t]{-12}{*}{\centering\arraybackslash Separate} & \multirow[t]{-4}{*}{\centering\arraybackslash Partial P3} & 0.25 & 0.00 & 0.00 & 1.00 & 0.00 & 0.00 & 0.00 & 0.00 & 0.00 & 0.00\\
\bottomrule
\end{tabular}
\end{table}

\begin{figure}[htbp]
    \centering
    \includegraphics[width=\textwidth]{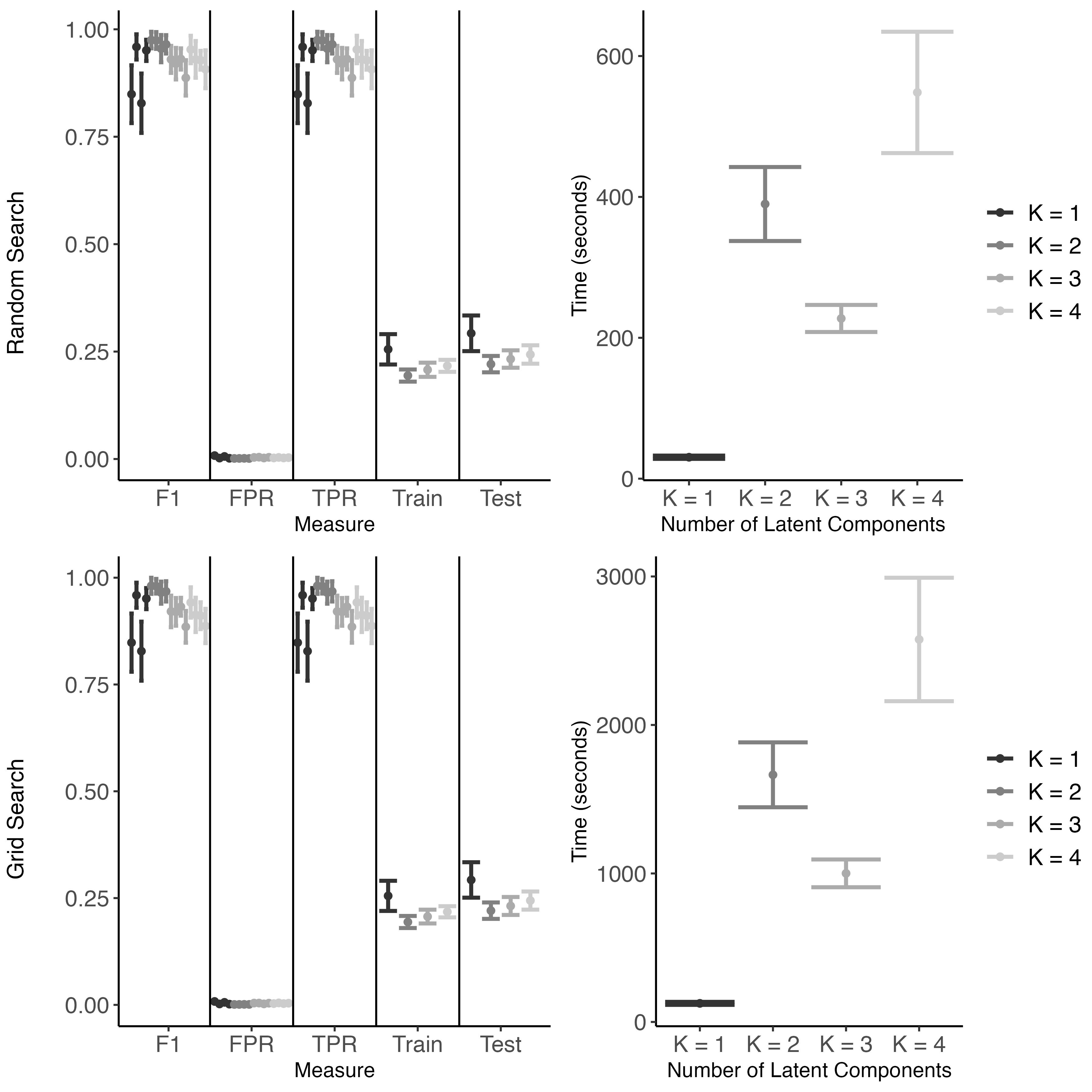}
    \caption{Sensitivity to selected K for Full Overlap Scenario. All results are for the P2 = [1000, 1500] variable dimension setting and are summarized over 20 simulated data sets.}\label{fig_Ksens_full}
\end{figure}

\begin{figure}[htbp]
    \centering
    \includegraphics[width=\textwidth]{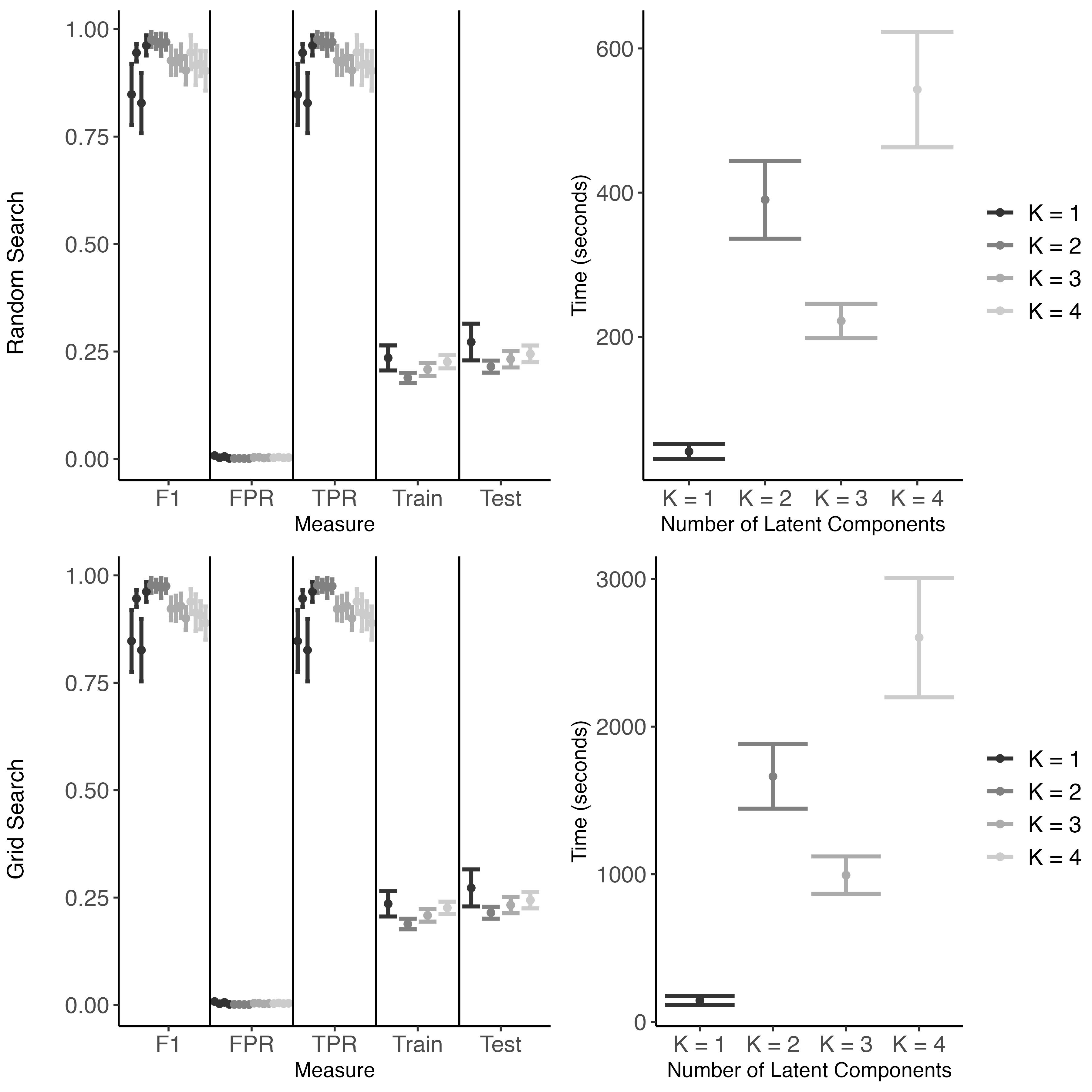}
    \caption{Sensitivity to selected K for Partial Overlap Scenario. All results are for the P2 = [1000, 1500] variable dimension setting and are summarized over 20 simulated data sets.}\label{fig_Ksens_partial}
\end{figure}

We also performed sensitivity analyses to determine how the results of the simulations would change based on different values of $K$. Figure \ref{fig_Ksens_full} and \ref{fig_Ksens_partial} show the variable selection performance, MSEs, and computation times using $K$ = 1, 2, 3, and 4 for the P2 variable dimension for the Full and Partial Overlap scenarios respectively. The results show that HIP is generally robust to a misspecification of $K$, but erring on the side of one too many components seems to be preferable to using one too few components.

% updated
\subsection{Tuning Parameter Selection}
As noted in Section 3.3 in the main text, there are two tuning parameters in HIP, $\lambda_\xi$ and $\lambda_g$. The hyperparameter space can be defined as all combinations of these two parameters. In particular, we create a grid with $a$ steps between $(0,\lambda_{max}]$ for both $\lambda_G$ and $\lambda_\xi$. Ideally, one would search over the entire grid space, but this is computationally expensive. Instead, we  use ideas in \cite{bergstra2012random} and randomly select $20\%$ of the $a^2$ grid values from the hyperparameter space to search over.

In the simulations, we set the range for $\lambda_\xi$ and $\lambda_g$ to $(0, 2]$ as this seemed to perform well for all variable dimensions tried. We set the number of steps to 8 which results in a grid of 64 points. HIP (Grid) searches over all 64 of these combinations, and HIP (Random) selects 13 of these combinations to search. Based on simulations, we recommend using the random search as results are similar between the two approaches, but the random search is computationally faster. 

We also implemented cross-validation as an alternative to the BIC, but the cross-validation is more computationally expensive and performed similarly to BIC. This cross-validation approach is available to users in the code.

%----------------------------------------------------
\section{Additional Details for Simulation Results}\label{appendix_sim}

\begin{figure}[htbp]
    \centering
    \includegraphics[width=0.85\textwidth]{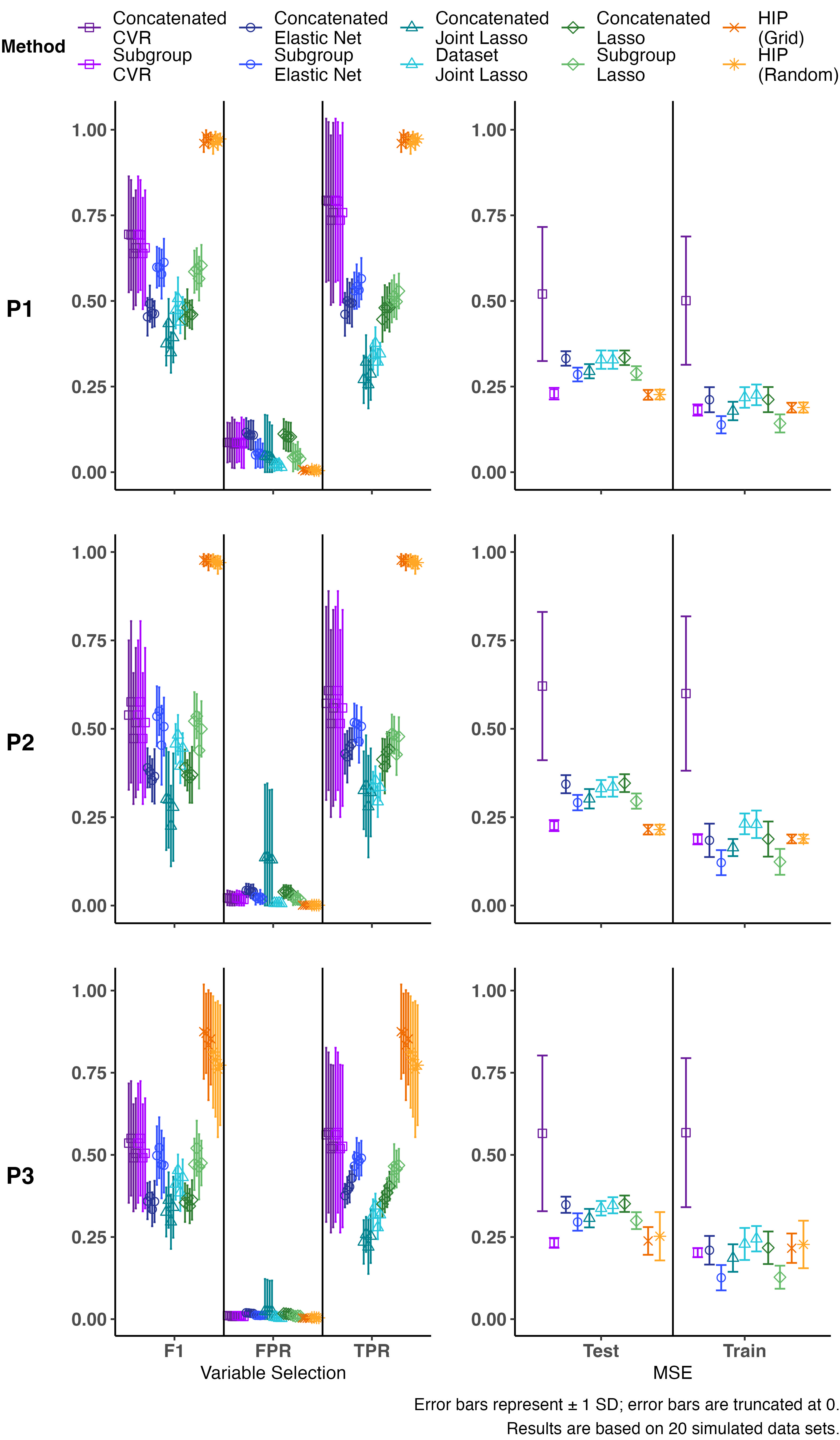}
    \caption{Results for Continuous Outcome, Partial Overlap Scenario. The first row corresponds to P1 ($p_1$ = 300, $p_2$ = 350), the second to P2 ($p_1$ = 1000, $p_2$ = 1500), and the third to P3 ($p_1$ = 2000, $p_2$ = 3000). For all settings, $n_1 = 250$ and $n_2 = 260$. The right column is test mean squared error (MSE), so a lower value indicates better performance. All results are based on 20 iterations.}\label{fig_performance_partial}
\end{figure}

Here we report additional details on our main simulation results. The results for Partial Overlap setting are similar to the Full Overlap setting (Figure \ref{fig_performance_partial}).  Again, the grid and random searches for the HIP provide similar results. There is a more distinct advantage to using our proposed method as CVR performs more poorly than in the Full Overlap scenario with a lower F1 and lower TPR. The other methods have a slightly lower F1 score than CVR. The Elastic Net, Lasso, and Joint Lasso have lower TPR rates, so they are missing important variables even  when they are run for the subgroups separately. The Joint Lasso has the highest FPR particularly in the P2 and P3 dimensions meaning that it is selecting a lot of variables that are not important. In terms of prediction, HIP (Grid), HIP (Random), and Subgroup CVR have lower Test MSEs than the other methods. Notably, Concatenated CVR has much a higher mean Test MSE and more variability than the other methods which highlights the importance of accounting for subgroup heterogeneity when it exists. The proposed method improves both variable selection and prediction. 

% Partial Overlap Performance

% Full Overlap Time
\begin{figure}[htbp]
    \centering
    \includegraphics[width=0.9\textwidth]{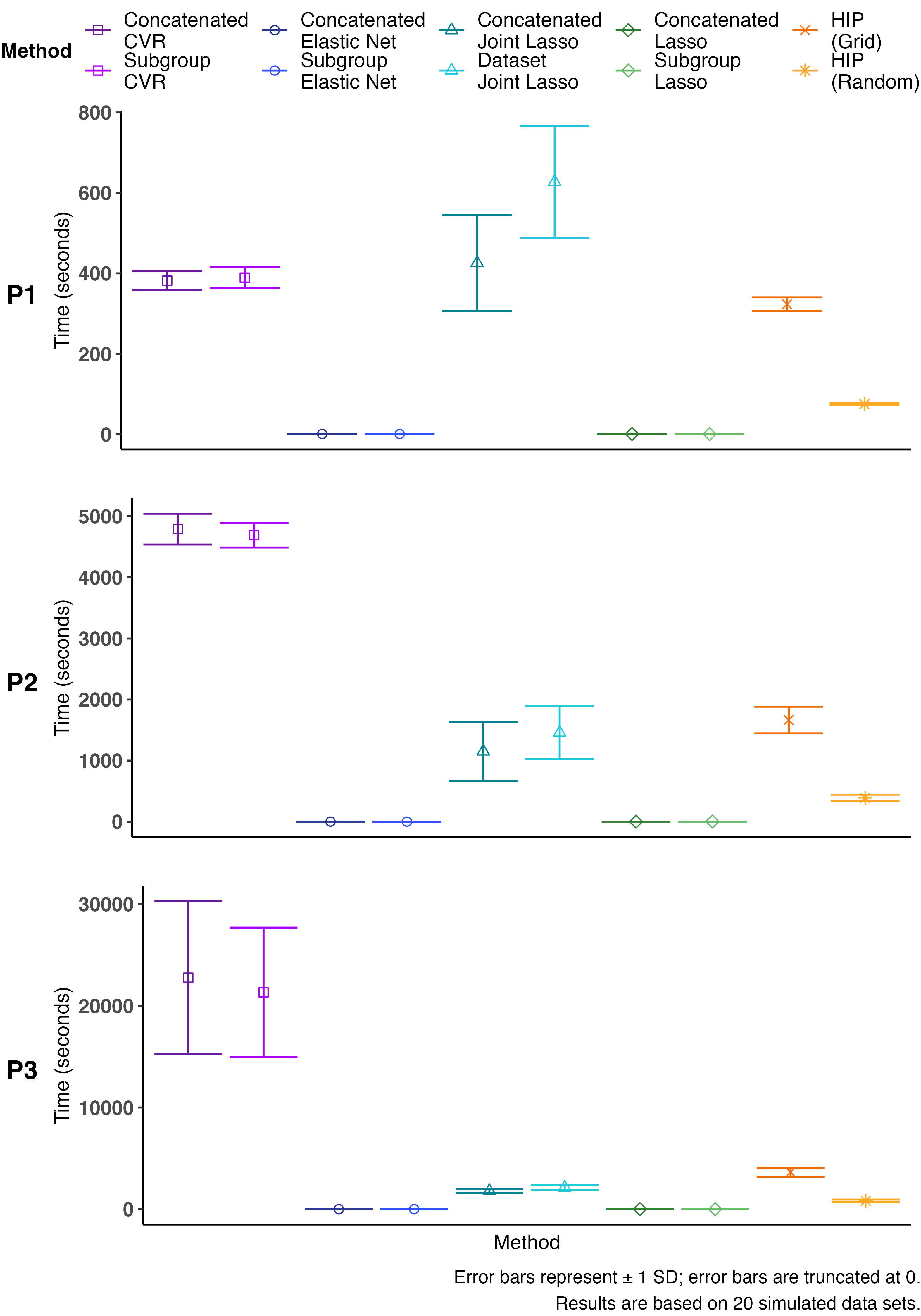}
    \caption{Computation Time for Continuous Outcome, Full Overlap Scenario. The Lasso and Elastic Net are the fastest methods across all parameter settings followed by the proposed method with random search. In the P1 setting, CVR and HIP (Grid) perform similarly, but in the P2 and P3 settings, the time required by CVR increases dramatically.}\label{fig_time_full}
\end{figure}

% Partial Overlap Time
\begin{figure}[htbp]
    \centering
    \includegraphics[width=0.9\textwidth]{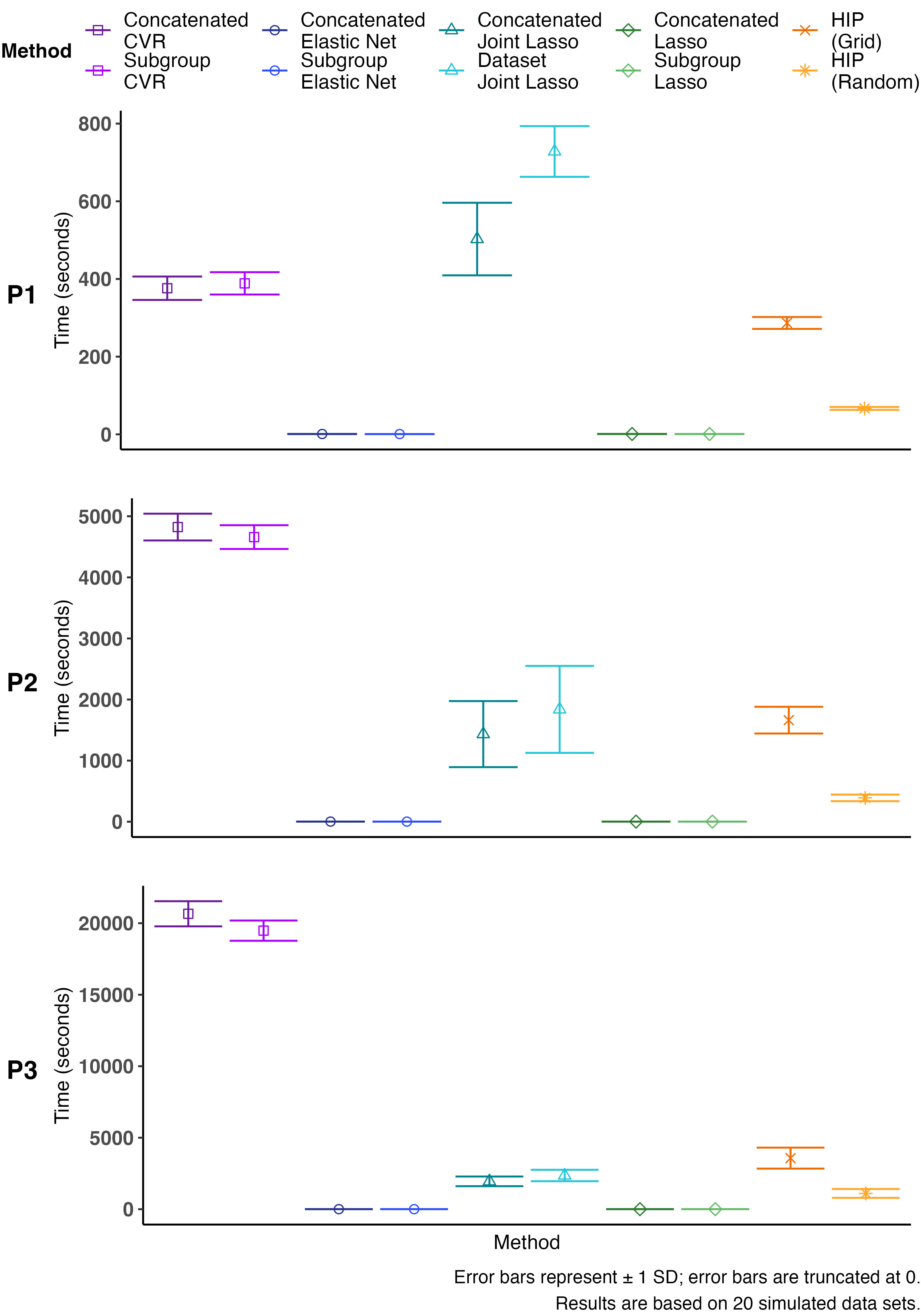}
    \caption{Computation Time for Continuous Outcome, Partial Overlap Scenario. The Lasso and Elastic Net are the fastest methods across all parameter settings followed by the proposed method with random search. In the P1 setting, CVR and HIP (Grid) perform similarly, but in the P2 and P3 settings, the time required by CVR increases dramatically.}\label{fig_time_partial}
\end{figure}

%----------------------------------------------------
\section{Additional Real Data Analysis Results}\label{appendix_rda}

\begin{figure}[htbp]
    \centering
    \includegraphics[width=\textwidth]{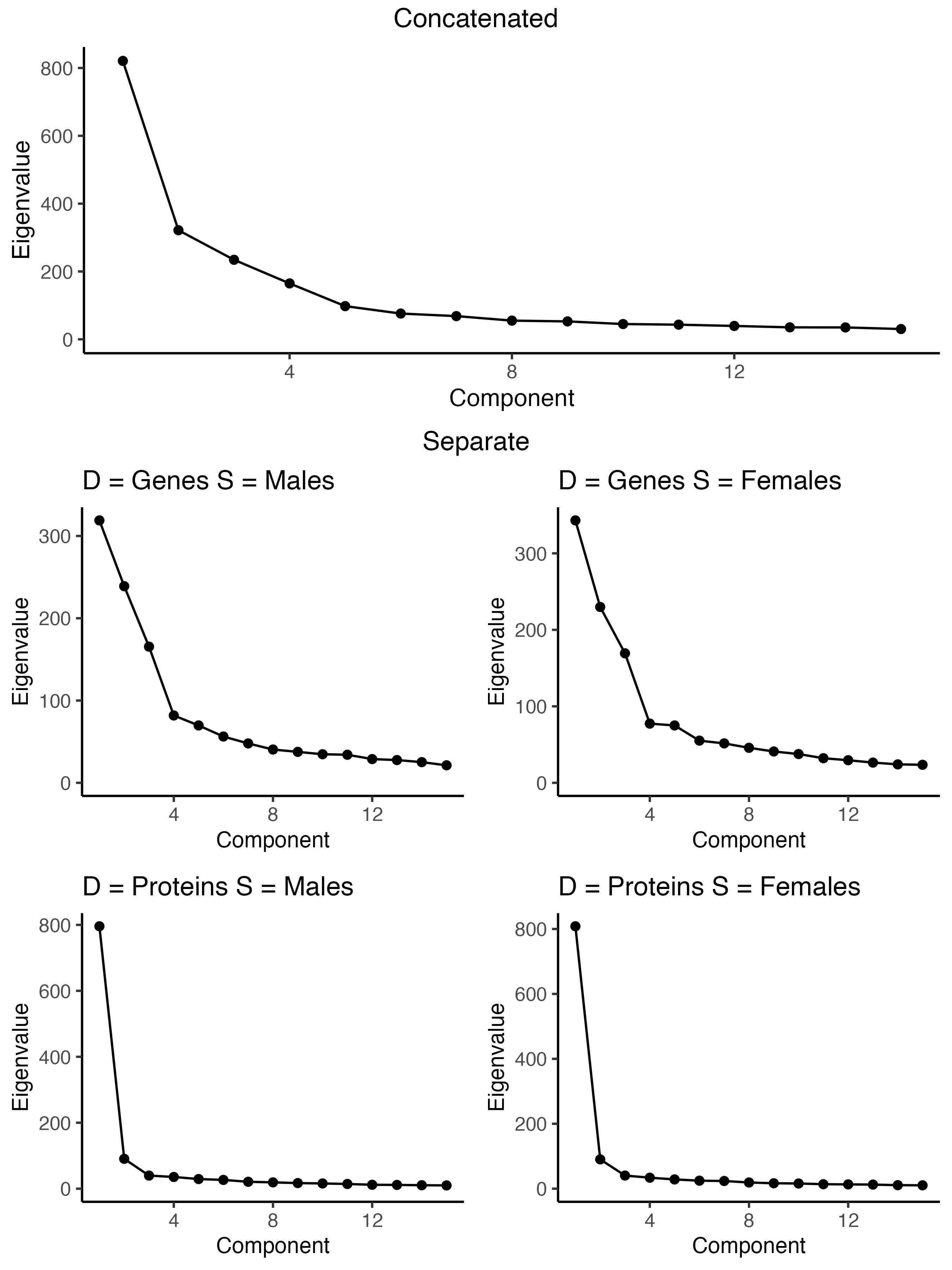}
    \caption{Scree plot of concatenated data matrix and each $\bm X^{d,s}$ for COPDGene data application. The data included 5000 genes and 2000 proteins that were selected from the unsupervised filtering.}\label{fig_RDA_scree}
\end{figure}

Here we report some additional findings from the data application using the COPDGene data. Figure \ref{fig_RDA_scree} shows the scree plot for the concatenated data after each $X^{d,s}$ was standardized by subgroup and the separate $X^{d,s}$.  The concatenated data plot seems to suggest a larger K than the separate plots. The separate protein data sets suggest a smaller $K$ than the gene data sets.

% Scree plots to select K

\begin{figure}[htbp]
    \centering
    \includegraphics[width=\textwidth]{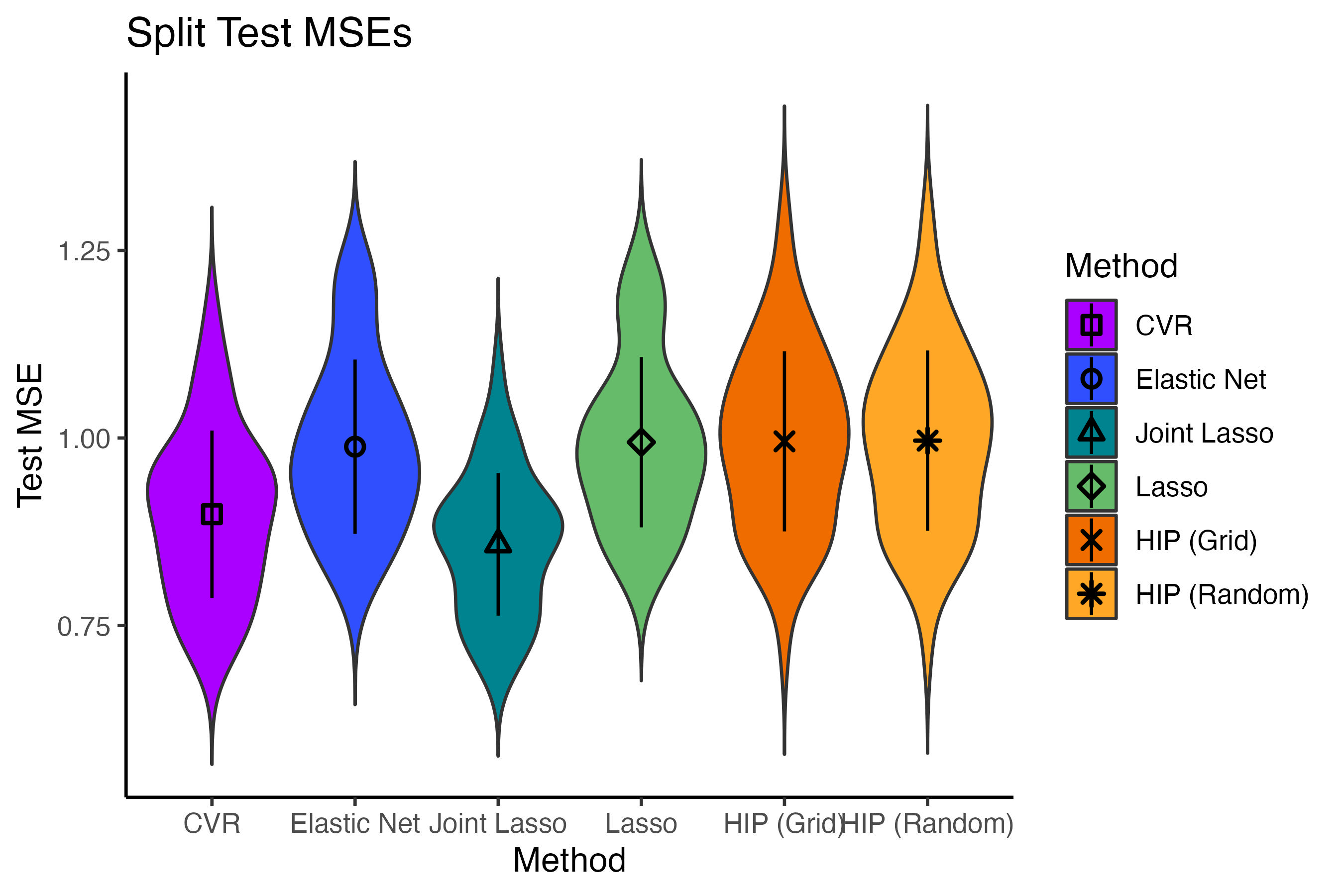}
    \caption{Distribution of test MSEs from fifty data splits}\label{fig_RDA_mses}
\end{figure}

\begin{figure}[htbp]
    \centering
    \includegraphics[width=\textwidth]{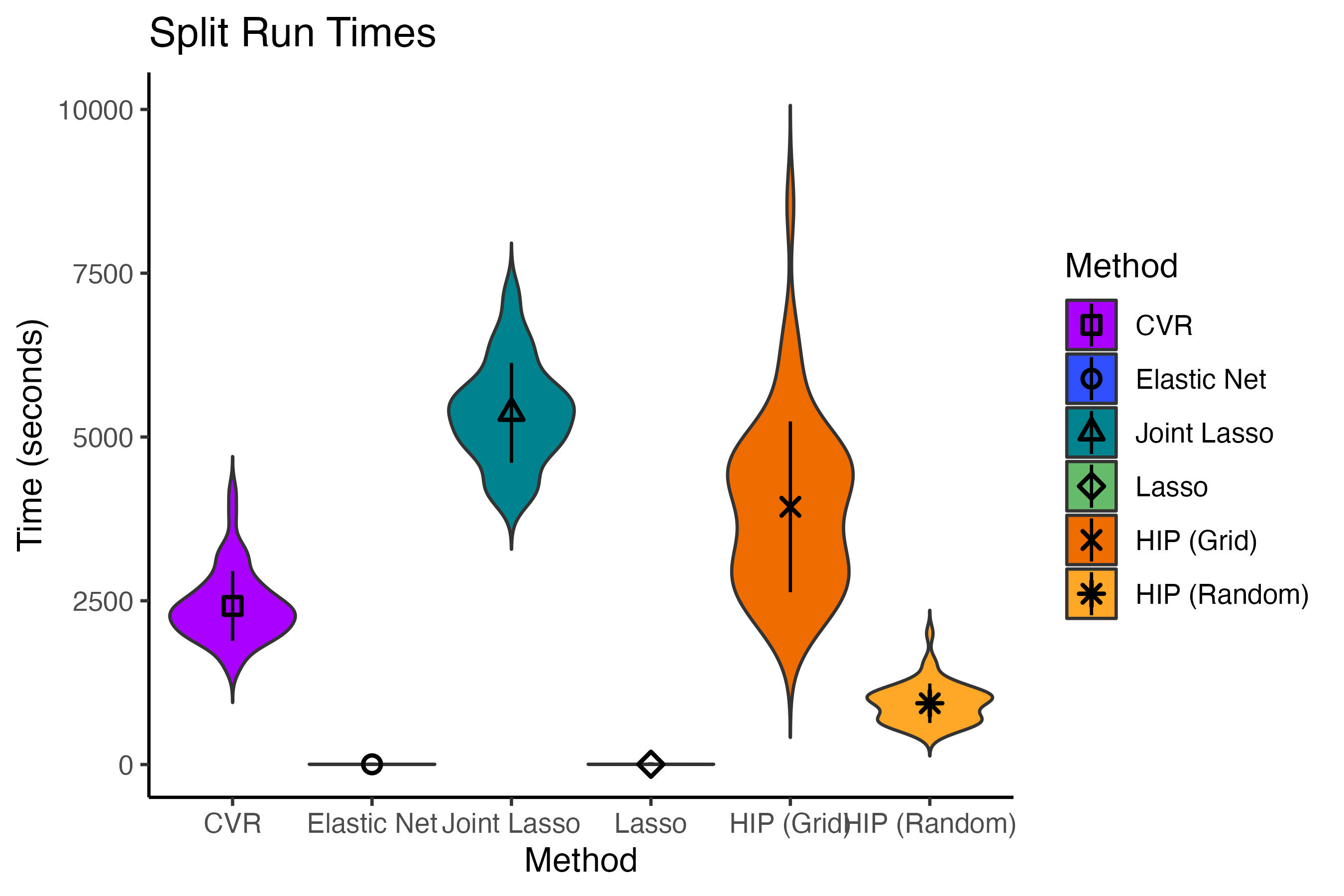}
    \caption{Distribution of run times from fifty data splits}\label{fig_RDA_times}
\end{figure}

Figure \ref{fig_RDA_mses} and Figure \ref{fig_RDA_times} show violin plots of the Test Mean Squared Errors (MSE) and run times from the fifty data splits for each method respectively. These MSEs were calculated with the variables selected in each split; the number of variables varied across the methods (Table \ref{tab_RDA_split_selected}). HIP (Grid) and HIP (Random) selects $N_{top} = 75$ genes and 25 proteins based on the estimates in $B^{d,s}$. Lasso and Elastic Net selected the next fewest variables followed by CVR. The Joint Lasso selected the most variables out of all of the methods applied to the data. While CVR and Joint Lasso have the smallest errors, they also use many more variables to do so. HIP (Grid) and HIP (Random) are able to produce comparable prediction errors with a much smaller number of variables.

% RDA MSEs

% RDA Times

% Number of variables used in predictions
\begin{table}[htbp]
\caption{Number of variables selected by each method from 50 data splits. Values are Mean (SD).}\label{tab_RDA_split_selected}%
\begin{tabular}{@{}lcccc@{}}
\toprule
\multicolumn{1}{c}{ } & \multicolumn{2}{c}{Genes} & \multicolumn{2}{c}{Proteins} \\
\cmidrule(l{3pt}r{3pt}){2-3} \cmidrule(l{3pt}r{3pt}){4-5}
Method & Males & Females & Males & Females\\
\midrule
CVR & 260.6 (39) & 194.9 (72) & 153 (26.3) & 153 (26.3)\\
Elastic Net & 175.9 (28) & 107.8 (34.8) & 66 (9.4) & 66 (9.4)\\
HIP (Grid) & 75 (0) & 75 (0) & 25 (0) & 25 (0)\\
HIP (Random) & 75 (0) & 75 (0) & 25 (0) & 25 (0)\\
Joint Lasso & 781.1 (62.8) & 777.3 (61.8) & 256.5 (81.5) & 256.5 (81.5)\\
Lasso & 167.9 (28.3) & 94.3 (36.2) & 62.8 (8.5) & 62.8 (8.5)\\
\bottomrule
\end{tabular}
\end{table}

Table \ref{tab_RDA_top_vars} shows the number of ``stable" genes and proteins (as defined in Section 5.2 of the main document) selected by each method and how many overlapped between males and females. We expect the numbers selected by each method to be close because we are selecting the top 1\% of genes and proteins from the 5000 genes and 2000 proteins that passed unsupervised filtering. Variation in the number selected is due to ties in the ranking of variables based on the number of splits in which the variable entered the model and the proportion of splits in which the variable was selected. Table \ref{tab_RDA_grid_v_rand} shows the number of variables selected by HIP (Grid) and  HIP (Random) and how many of the selected variables overlapped. The grid and random searches result in very similar sets of ``stable" genes and proteins which also supports the use of the random search.

% number of variables selected as stable for all methods
\begin{table}[htbp]
\caption{Number of Stable Variables Selected by Each Method. The top 1\% (of the 5000 genes and 2000 proteins retained from unsupervised filtering) most frequently selected genes and proteins across 50 random splits of the data are considered 'selected'.}\label{tab_RDA_top_vars}%
\begin{tabular}{@{}lcccccc@{}}
\toprule
\multicolumn{1}{c}{ } & \multicolumn{3}{c}{Genes} & \multicolumn{3}{c}{Proteins} \\
\cmidrule(l{3pt}r{3pt}){2-4} \cmidrule(l{3pt}r{3pt}){5-7}
Method & Males & Females & Common & Males & Females & Common\\
\midrule
HIP (Random) & 50 & 50 & 46 & 22 & 20 & 9\\
HIP (Grid) & 50 & 50 & 47 & 20 & 20 & 8\\
Joint Lasso & 64 & 63 & 48 & 34 & 38 & 29\\
CVR & 50 & 50 & 15 & 21 & 23 & 12\\
Lasso & 50 & 50 & 4 & 20 & 20 & 5\\
Elastic Net & 52 & 50 & 6 & 20 & 20 & 6\\
\bottomrule
\end{tabular}
\end{table}

% Compare variables selected by grid and random searches
\begin{table}[htbp]
\caption{Variables Selected by Random and Grid Searches Using HIP. There is strong agreement in the variables selected using either search method.}\label{tab_RDA_grid_v_rand}%
\begin{tabular}{@{}ccccc@{}}
\toprule
Data & Subtype & Random & Grid & Overlap between Random and Grid\\
\midrule
 & Males & 50 & 50 & 50\\
\cmidrule{2-5}
 & Females & 50 & 50 & 48\\
\cmidrule{2-5}
\multirow[t]{-3}{*}{\centering\arraybackslash Genes} & Common & 46 & 47 & 45\\
\cmidrule{1-5}
 & Males & 22 & 20 & 20\\
\cmidrule{2-5}
 & Females & 20 & 20 & 20\\
\cmidrule{2-5}
\multirow[t]{-3}{*}{\centering\arraybackslash Proteins} & Common & 9 & 8 & 8\\
\bottomrule
\end{tabular}
\end{table}

\begin{figure}[htbp]
    \centering
    \includegraphics[width=\textwidth]{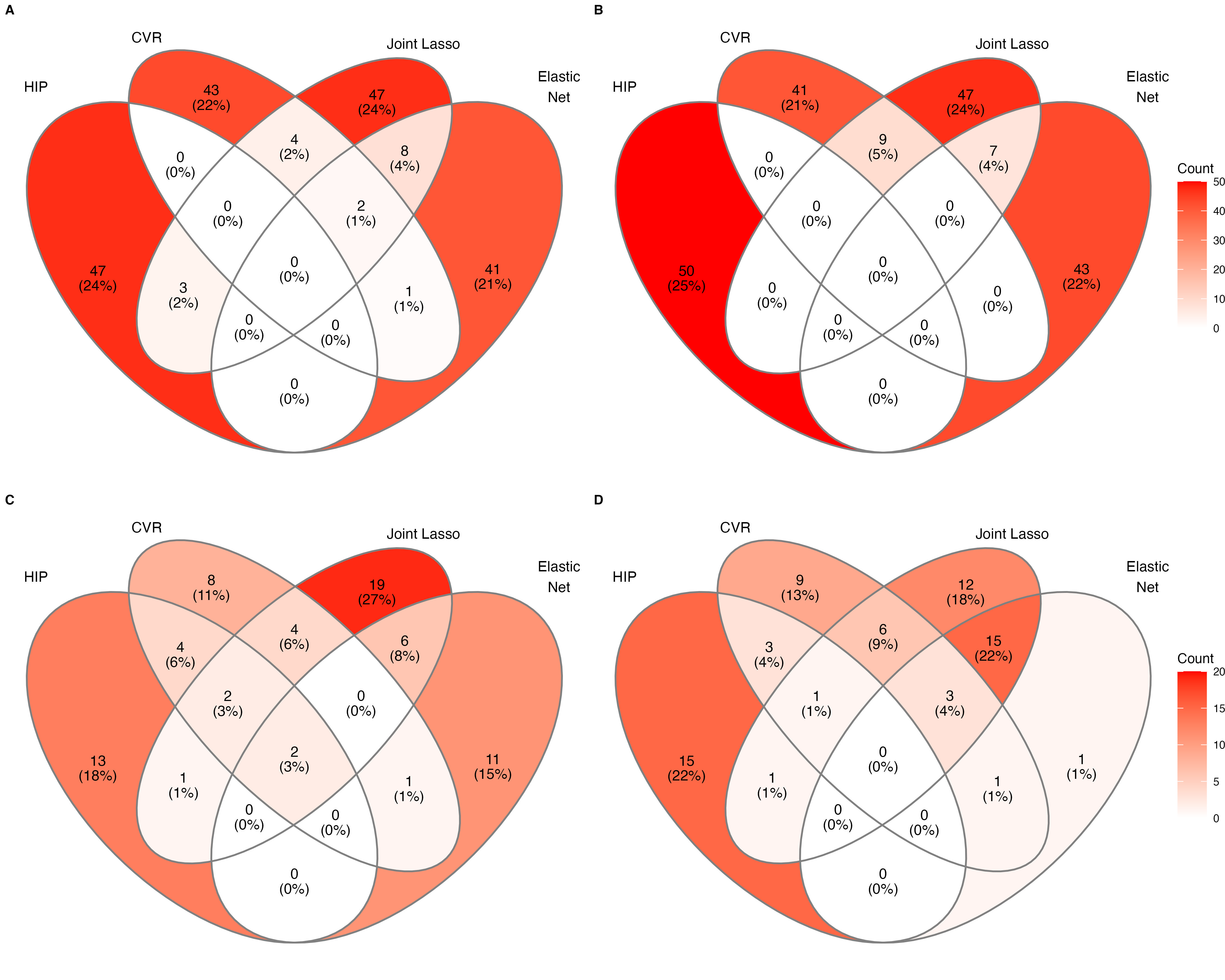}
    \caption{A) Overlap in genes identified for males; B) Overlap in genes identified for females; C) Overlap in proteins identified for males; D) Overlap in proteins identified for females.}\label{fig_RDA_venn}
\end{figure}

Figure \ref{fig_RDA_venn} shows the overlap of ``stable" genes and proteins selected by HIP (Random), CVR, Joint Lasso, and Elastic Net.  HIP (Grid) and Lasso were not included to prevent over-cluttering as the variables selected by these two methods were similar to HIP (Random) and Elastic Net respectively. There are very few overlaps in the ``stable" variables selected by these methods particularly for the genes. The  genes selected for females by HIP (Random) did not overlap at all with the other methods, and the genes selected for males by HIP (Random) only shared 3 genes with Joint Lasso and 0 genes with the other methods.

% Venn diagram of overlaps

Table \ref{tab_RDA_weights_prot} lists the proteins and Tables \ref{tab_RDA_weights_geneM} and \ref{tab_RDA_weights_geneF} list the genes selected by HIP (Random) for males and females in addition to the average weights calculated from the $\bm{B}^{d,s}$ matrices.  

% selected genes/proteins and weights
% Genes - Males
\begin{table}[htbp]
\begin{scriptsize}
\caption{List of Genes Selected by HIP (Random) for Males in COPD Application. Weights are an average calculated from the estimated $\bm B^{d,s}$ matrices. Higher values suggest the variable likely contributes more to association and prediction.}\label{tab_RDA_weights_geneM}%
\begin{tabular*}{\textwidth}{@{\extracolsep{\fill}}p{1in}p{2in}p{0.5in}p{0.5in}}
\toprule
ID & Name & Symbol & Weight\\
\midrule
ENSG00000159346.13 & adiponectin receptor 1 & ADIPOR1 & 1.121\\
ENSG00000124098.10 & family with sequence similarity 210 member B & FAM210B & 1.112\\
ENSG00000130830.15 & MAGUK p55 scaffold protein 1 & MPP1 & 1.107\\
ENSG00000198876.13 & DDB1 and CUL4 associated factor 12 & DCAF12 & 1.102\\
ENSG00000099804.9 & cell division cycle 34, ubiqiutin conjugating enzyme & CDC34 & 1.101\\
ENSG00000137198.10 & guanosine monophosphate reductase & GMPR & 1.100\\
ENSG00000133193.12 & family with sequence similarity 104 member A & FAM104A & 1.100\\
ENSG00000171552.14 & BCL2 like 1 & BCL2L1 & 1.098\\
ENSG00000147454.14 & solute carrier family 25 member 37 & SLC25A37 & 1.097\\
ENSG00000167671.12 & UBX domain protein 6 & UBXN6 & 1.096\\
ENSG00000119950.21 & MAX interactor 1, dimerization protein & MXI1 & 1.095\\
ENSG00000196961.13 & adaptor related protein complex 2 subunit alpha 1 & AP2A1 & 1.093\\
ENSG00000143321.19 & heparin binding growth factor & HDGF & 1.092\\
ENSG00000168785.8 & tetraspanin 5 & TSPAN5 & 1.091\\
ENSG00000166947.15 & erythrocyte membrane protein band 4.2 & EPB42 & 1.089\\
ENSG00000143416.21 & selenium binding protein 1 & SELENBP1 & 1.088\\
ENSG00000022840.16 & ring finger protein 10 & RNF10 & 1.085\\
ENSG00000101782.15 & RIO kinase 3 & RIOK3 & 1.085\\
ENSG00000158856.18 & dematin actin binding protein & DMTN & 1.085\\
ENSG00000105701.16 & FKBP prolyl isomerase 8 & FKBP8 & 1.085\\
ENSG00000133606.11 & makorin ring finger protein 1 & MKRN1 & 1.084\\
ENSG00000066926.12 & ferrochelatase & FECH & 1.084\\
ENSG00000183508.5 & terminal nucleotidyltransferase 5C & TENT5C & 1.081\\
ENSG00000100614.18 & protein phosphatase, Mg2+/Mn2+ dependent 1A & PPM1A & 1.081\\
ENSG00000182512.5 & glutaredoxin 5 & GLRX5 & 1.079\\
ENSG00000079308.19 & tensin 1 & TNS1 & 1.077\\
ENSG00000070182.21 & spectrin beta, erythrocytic & SPTB & 1.077\\
ENSG00000013306.16 & solute carrier family 25 member 39 & SLC25A39 & 1.074\\
ENSG00000162722.9 & tripartite motif containing 58 & TRIM58 & 1.074\\
ENSG00000167705.12 & Rab interacting lysosomal protein & RILP & 1.074\\
ENSG00000130821.17 & solute carrier family 6 member 8 & SLC6A8 & 1.071\\
ENSG00000103342.13 & G1 to S phase transition 1 & GSPT1 & 1.070\\
ENSG00000082146.13 & STE20 related adaptor beta & STRADB & 1.069\\
ENSG00000143774.17 & guanylate kinase 1 & GUK1 & 1.069\\
ENSG00000034713.8 & GABA type A receptor associated protein like 2 & GABARAPL2 & 1.061\\
ENSG00000102145.15 & GATA binding protein 1 & GATA1 & 1.061\\
ENSG00000063854.13 & hydroxyacylglutathione hydrolase & HAGH & 1.060\\
ENSG00000132819.17 & RNA binding motif protein 38 & RBM38 & 1.059\\
ENSG00000004939.15 & solute carrier family 4 member 1 (Diego blood group) & SLC4A1 & 1.059\\
ENSG00000060138.13 & Y-box binding protein 3 & YBX3 & 1.057\\
ENSG00000105281.12 & solute carrier family 1 member 5 & SLC1A5 & 1.055\\
ENSG00000181788.4 & siah E3 ubiquitin protein ligase 2 & SIAH2 & 1.051\\
ENSG00000136732.16 & glycophorin C (Gerbich blood group) & GYPC & 1.048\\
ENSG00000137216.19 & transmembrane protein 63B & TMEM63B & 1.047\\
ENSG00000047597.7 & X-linked Kx blood group antigen, Kell and VPS13A binding protein & XK & 1.046\\
ENSG00000153574.9 & ribose 5-phosphate isomerase A & RPIA & 1.045\\
ENSG00000100325.15 & activating signal cointegrator 1 complex subunit 2 & ASCC2 & 1.040\\
ENSG00000205639.12 & major facilitator superfamily domain containing 2B & MFSD2B & 1.040\\
ENSG00000204613.11 & tripartite motif containing 10 & TRIM10 & 1.030\\
ENSG00000129472.15 & RAB2B, member RAS oncogene family & RAB2B & 1.014\\
\bottomrule
\end{tabular*}
\end{scriptsize}
\end{table}

% Genes - Females 
\begin{table}[htbp]
\begin{scriptsize}
\caption{List of Genes Selected by HIP (Random) for Females in COPD Application. Weights are an average calculated from the estimated $\bm B^{d,s}$ matrices. Higher values suggest the variable likely contributes more to association and prediction.}\label{tab_RDA_weights_geneF}%
\begin{tabular*}{\textwidth}{@{\extracolsep{\fill}}p{1in}p{2in}p{0.5in}p{0.5in}}
\toprule%
ID & Name & Symbol & Weight\\
\midrule
ENSG00000171552.14 & BCL2 like 1 & BCL2L1 & 1.162\\
ENSG00000159346.13 & adiponectin receptor 1 & ADIPOR1 & 1.155\\
ENSG00000198876.13 & DDB1 and CUL4 associated factor 12 & DCAF12 & 1.152\\
ENSG00000130821.17 & solute carrier family 6 member 8 & SLC6A8 & 1.145\\
ENSG00000124098.10 & family with sequence similarity 210 member B & FAM210B & 1.140\\
ENSG00000130830.15 & MAGUK p55 scaffold protein 1 & MPP1 & 1.135\\
ENSG00000137198.10 & guanosine monophosphate reductase & GMPR & 1.129\\
ENSG00000073792.16 & insulin like growth factor 2 mRNA binding protein 2 & IGF2BP2 & 1.128\\
ENSG00000082146.13 & STE20 related adaptor beta & STRADB & 1.127\\
ENSG00000060138.13 & Y-box binding protein 3 & YBX3 & 1.127\\
ENSG00000166947.15 & erythrocyte membrane protein band 4.2 & EPB42 & 1.125\\
ENSG00000133606.11 & makorin ring finger protein 1 & MKRN1 & 1.125\\
ENSG00000119950.21 & MAX interactor 1, dimerization protein & MXI1 & 1.122\\
ENSG00000158856.18 & dematin actin binding protein & DMTN & 1.122\\
ENSG00000168785.8 & tetraspanin 5 & TSPAN5 & 1.120\\
ENSG00000143416.21 & selenium binding protein 1 & SELENBP1 & 1.119\\
ENSG00000066926.12 & ferrochelatase & FECH & 1.117\\
ENSG00000183508.5 & terminal nucleotidyltransferase 5C & TENT5C & 1.116\\
ENSG00000013306.16 & solute carrier family 25 member 39 & SLC25A39 & 1.116\\
ENSG00000133193.12 & family with sequence similarity 104 member A & FAM104A & 1.115\\
ENSG00000099804.9 & cell division cycle 34, ubiqiutin conjugating enzyme & CDC34 & 1.114\\
ENSG00000162722.9 & tripartite motif containing 58 & TRIM58 & 1.113\\
ENSG00000105701.16 & FKBP prolyl isomerase 8 & FKBP8 & 1.111\\
ENSG00000022840.16 & ring finger protein 10 & RNF10 & 1.111\\
ENSG00000143321.19 & heparin binding growth factor & HDGF & 1.110\\
ENSG00000182512.5 & glutaredoxin 5 & GLRX5 & 1.110\\
ENSG00000167671.12 & UBX domain protein 6 & UBXN6 & 1.110\\
ENSG00000196961.13 & adaptor related protein complex 2 subunit alpha 1 & AP2A1 & 1.108\\
ENSG00000147454.14 & solute carrier family 25 member 37 & SLC25A37 & 1.108\\
ENSG00000070182.21 & spectrin beta, erythrocytic & SPTB & 1.108\\
ENSG00000004939.15 & solute carrier family 4 member 1 (Diego blood group) & SLC4A1 & 1.107\\
ENSG00000102145.15 & GATA binding protein 1 & GATA1 & 1.103\\
ENSG00000103342.13 & G1 to S phase transition 1 & GSPT1 & 1.101\\
ENSG00000079308.19 & tensin 1 & TNS1 & 1.101\\
ENSG00000132819.17 & RNA binding motif protein 38 & RBM38 & 1.100\\
ENSG00000100614.18 & protein phosphatase, Mg2+/Mn2+ dependent 1A & PPM1A & 1.098\\
ENSG00000101782.15 & RIO kinase 3 & RIOK3 & 1.095\\
ENSG00000137216.19 & transmembrane protein 63B & TMEM63B & 1.090\\
ENSG00000167705.12 & Rab interacting lysosomal protein & RILP & 1.086\\
ENSG00000063854.13 & hydroxyacylglutathione hydrolase & HAGH & 1.081\\
ENSG00000143774.17 & guanylate kinase 1 & GUK1 & 1.075\\
ENSG00000047597.7 & X-linked Kx blood group antigen, Kell and VPS13A binding protein & XK & 1.073\\
ENSG00000129472.15 & RAB2B, member RAS oncogene family & RAB2B & 1.072\\
ENSG00000186591.13 & ubiquitin conjugating enzyme E2 H & UBE2H & 1.070\\
ENSG00000153574.9 & ribose 5-phosphate isomerase A & RPIA & 1.069\\
ENSG00000105281.12 & solute carrier family 1 member 5 & SLC1A5 & 1.066\\
ENSG00000162367.11 & TAL bHLH transcription factor 1, erythroid differentiation factor & TAL1 & 1.064\\
ENSG00000181788.4 & siah E3 ubiquitin protein ligase 2 & SIAH2 & 1.064\\
ENSG00000158578.21 & 5'-aminolevulinate synthase 2 & ALAS2 & 1.040\\
ENSG00000204613.11 & tripartite motif containing 10 & TRIM10 & 1.039\\
\bottomrule
\end{tabular*}
\end{scriptsize}
\end{table}

% Proteins - Males and Females
\begin{table}[htbp]
\begin{scriptsize}
\caption{List of Proteins Selected by HIP (Random) in COPD Application. Weights are an average calculated from the estimated $\bm B^{d,s}$ matrices. Higher values suggest the variable likely contributes more to association and prediction.}\label{tab_RDA_weights_prot}%
\begin{tabular*}{\textwidth}{@{\extracolsep{\fill}}p{0.5in}p{0.5in}p{2in}p{0.5in}p{0.5in}}
\toprule
Subgroup & ID & Name & Symbol & Weight\\
\midrule

 & 55666 & Nuclear protein localization protein 4 homolog & NPLOC4 & 1.310\\

 & 51324 & Maspardin & SPG21 & 1.273\\

 & 5196 & Platelet factor 4 & PF4 & 1.033\\

 & 2814 & Platelet glycoprotein V & GP5 & 1.032\\

 & 5473 & Beta-thromboglobulin & PPBP & 1.030\\

 & 627 & Brain-derived neurotrophic factor & BDNF & 0.962\\

 & 115761 & ADP-ribosylation factor-like protein 11 & ARL11 & 0.958\\

 & 5473 & Platelet basic protein & PPBP & 0.931\\

 & 5534 & Calcineurin subunit B type 1 & PPP3R1 & 0.925\\

 & 5777 & Tyrosine-protein phosphatase non-receptor type 6 & PTPN6 & 0.827\\

 & 10814 & Complexin-2 & CPLX2 & 0.797\\

 & 340205 & Trem-like transcript 1 protein:immunoreceptor tyrosine-based inhibition motif & TREML1 & 0.784\\

 & 81567 & Thioredoxin domain-containing protein 5 & TXNDC5 & 0.734\\

 & 4879 & N-terminal pro-BNP & NPPB & 0.729\\

 & 3485 & Insulin-like growth factor-binding protein 2 & IGFBP2 & 0.715\\

 & 2056 & Erythropoietin & EPO & 0.664\\

 & 9370 & Adiponectin & ADIPOQ & 0.586\\

 & 3630 & Insulin & INS & 0.584\\

 & 3952 & Leptin & LEP & 0.579\\

 & 5069 & Pappalysin-1 & PAPPA & 0.574\\

 & 3934 & Neutrophil gelatinase-associated lipocalin & LCN2 & 0.516\\

\multirow[t]{-22}{*}{\centering\arraybackslash Males} & 6289 & Serum amyloid A-2 protein & SAA2 & 0.446\\

\cmidrule{1-5}

 & 60682 & Stromal membrane-associated protein 1 & SMAP1 & 1.192\\

 & 1032 & Cyclin-dependent kinase 4 inhibitor D & CDKN2D & 1.188\\

 & 5473 & Connective tissue-activating peptide III & PPBP & 1.157\\

 & 2814 & Platelet glycoprotein V & GP5 & 1.156\\

 & 1950 & Epidermal growth factor:Extracellular domain & EGF & 1.133\\

 & 627 & Brain-derived neurotrophic factor & BDNF & 1.129\\

 & 5473 & Neutrophil-activating peptide 2 & PPBP & 1.116\\

 & 5196 & Platelet factor 4 & PF4 & 1.087\\

 & 5473 & Beta-thromboglobulin & PPBP & 1.072\\

 & 5473 & Platelet basic protein & PPBP & 1.068\\

 & 7132 & Tumor necrosis factor receptor superfamily member 1A & TNFRSF1A & 0.972\\

 & 10406 & WAP four-disulfide core domain protein 2 & WFDC2 & 0.921\\

 & 51087 & Y-box-binding protein 2 & YBX2 & 0.915\\

 & 4904 & Nuclease-sensitive element-binding protein 1 & YBX1 & 0.802\\

 & 10814 & Complexin-2 & CPLX2 & 0.799\\

 & 11314 & CMRF35-like molecule 8 & CD300A & 0.796\\

 & 3485 & Insulin-like growth factor-binding protein 2 & IGFBP2 & 0.700\\

 & 1088 & Carcinoembryonic antigen-related cell adhesion molecule 8 & CEACAM8 & 0.670\\

 & 5069 & Pappalysin-1 & PAPPA & 0.651\\

\multirow[t]{-20}{*}{\centering\arraybackslash Females} & 9370 & Adiponectin & ADIPOQ & 0.619\\
\bottomrule
\end{tabular*}
\end{scriptsize}
\end{table}

\clearpage

\bibliographystyle{plain}
\bibliography{References.bib}